\documentclass[preprint2]{aastex63}
\pdfoutput=1 
\usepackage{amsmath,amstext}
\usepackage[T1]{fontenc}
\usepackage{apjfonts} 
\usepackage[figure,figure*]{hypcap}
\usepackage{color}
\usepackage{booktabs}
\usepackage{longtable}
\usepackage{multirow}
\usepackage{subfigure}
\usepackage{mathrsfs}

\newcommand{\unit}[1]{\ensuremath{\, \mathrm{#1}}}

\newcommand{\vsini}{$v \sin i$~}
\newcommand{\vsinip}{$v \sin i$}
\newcommand{\kep}{\emph{Kepler}~}
\newcommand{\kepp}{\emph{Kepler}}
\newcommand{\lightkurve}{\texttt{lightkurve}~}
\newcommand{\lightkurvep}{\texttt{lightkurve}}
\newcommand{\celerite}{\texttt{celerite}~}

\newcommand{\emcee}{\texttt{emcee}~}

\newcommand{\radvelp}{\texttt{radvel}}
\newcommand{\iprime}{$i^\prime$~}

\usepackage{hyperref}
\usepackage{breakurl}
\shorttitle{Persistent starspot signals on M dwarfs}
\shortauthors{Robertson et al. 2020}

\newcommand{\PSUAA}{Department of Astronomy \& Astrophysics, 525 Davey Laboratory, The Pennsylvania State University, University Park, PA, 16802, USA}
\newcommand{\PSUCEHW}{Center for Exoplanets and Habitable Worlds, 525 Davey Laboratory, The Pennsylvania State University, University Park, PA, 16802, USA}

\newcommand{\UA}{Steward Observatory, The University of Arizona, 933 N.\ Cherry Ave, Tucson, AZ 85721, USA}

\newcommand{\Caltech}{Department of Astronomy, California Institute of Technology, Pasadena, CA 91125, USA}

\newcommand{\Macquarie}{Department of Physics and Astronomy, Macquarie University, Balaclava Road, North Ryde, NSW 2109, Australia }
\newcommand{\NIST}{Time and Frequency Division, National Institute of Standards \& Technology, 325 Broadway, Boulder, CO 80305, USA}
\newcommand{\CUBoulder}{Department of Physics, 390 UCB, University of Colorado, Boulder, CO 80309, USA}
\newcommand{\JPL}{Jet Propulsion Laboratory, California Institute of Technology, 4800 Oak Grove Drive, Pasadena, California 91109}

\newcommand{\UCI}{Department of Physics \& Astronomy, The University of California, Irvine, Irvine, CA 92697, USA}
\newcommand{\Carleton}{Carleton College, One North College St., Northfield, MN 55057, USA}
\newcommand{\PSUICS}{Institute for Computational and Data Sciences, The Pennsylvania State University, University Park, PA, 16802, USA}

\newcommand{\Princeton}{Department of Astrophysical Sciences, Princeton University, 4 Ivy Lane, Princeton, NJ 08540, USA}
\newcommand{\UT}{Department of Astronomy and McDonald Observatory, University of Texas at Austin, 2515 Speedway, Austin, TX 78712, USA}

\newcommand{\CPSH}{Center for Planetary Systems Habitability, University of Texas at Austin, Austin TX 78712 USA}
\newcommand{\USAF}{Space Vehicles Directorate, Air Force Research Laboratory, 3550 Aberdeen Ave. SE, Kirtland AFB, NM 87117, USA}

\begin{document}
\title{Persistent starspot signals on M dwarfs: multi-wavelength Doppler observations with the Habitable-zone Planet Finder and Keck/HIRES}

\author[0000-0003-0149-9678]{Paul Robertson}
\affil{\UCI}

\author[0000-0001-7409-5688]{Gudmundur Stefansson}
\altaffiliation{Henry Norris Russell Fellow}
\affil{\Princeton}

\author[0000-0001-9596-7983]{Suvrath Mahadevan}
\affil{\PSUAA}
\affil{\PSUCEHW}

\author{Michael Endl}
\affil{\UT}

\author[0000-0001-9662-3496]{William~D.~Cochran}
\affil{\UT}
\affil{\CPSH}

\author[0000-0001-7708-2364]{Corey Beard}
\affil{\UCI}

\author[0000-0003-4384-7220]{Chad F.\ Bender}
\affil{\UA}

\author[0000-0002-2144-0764]{Scott A.\ Diddams}
\affil{\NIST}
\affil{\CUBoulder}

\author[0000-0001-6730-9115]{Nicholas Duong}
\affil{\UCI}

\author[0000-0001-6545-639X]{Eric B.\ Ford}
\affil{\PSUAA}
\affil{\PSUCEHW}
\affil{\PSUICS}

\author[0000-0002-0560-1433]{Connor Fredrick}
\affil{\NIST}
\affil{\CUBoulder}

\author[0000-0003-1312-9391]{Samuel Halverson}
\affil{\JPL}

\author[0000-0002-1664-3102]{Fred Hearty}
\affil{\PSUAA}
\affil{\PSUCEHW}

\author[0000-0002-5034-9476]{Rae Holcomb}
\affil{\UCI}

\author[0000-0003-1035-3232]{Lydia Juan}
\affil{\UCI}

\author[0000-0001-8401-4300]{Shubham Kanodia}
\affil{\PSUAA}
\affil{\PSUCEHW}

\author[0000-0001-8342-7736]{Jack Lubin}
\affil{\UCI}

\author[0000-0001-5000-1018]{Andrew~J.~Metcalf}
\affil{\USAF}
\affil{\NIST}
\affil{\CUBoulder}

\author[0000-0002-0048-2586]{Andrew Monson}
\affil{\PSUAA}
\affil{\PSUCEHW}

\author[0000-0001-8720-5612]{Joe P.\ Ninan}
\affil{\PSUAA}
\affil{\PSUCEHW}

\author[0000-0002-3411-8897]{Jonathan Palafoutas}
\affil{\UCI}

\author{Lawrence W.\ Ramsey}
\affil{\PSUAA}
\affil{\PSUCEHW}

\author[0000-0001-8127-5775]{Arpita Roy}
\altaffiliation{Robert A Millikan Postdoctoral Fellow}
\affil{\Caltech}

\author[0000-0002-4046-987X]{Christian Schwab}
\affil{\Macquarie}

\author[0000-0002-4788-8858]{Ryan C. Terrien}
\affil{\Carleton}

\author[0000-0001-6160-5888]{Jason T.\ Wright}
\affil{\PSUAA}
\affil{\PSUCEHW}

\email{paul.robertson@uci.edu}

\begin{abstract}
Young, rapidly-rotating M dwarfs exhibit prominent starspots, which create quasiperiodic signals in their photometric and Doppler spectroscopic measurements.  The periodic Doppler signals can mimic radial velocity (RV) changes expected from orbiting exoplanets.  Exoplanets can be distinguished from activity-induced false positives by the chromaticity and long-term incoherence of starspot signals, but these qualities are poorly constrained for fully-convective M stars.  Coherent photometric starspot signals on M dwarfs may persist for hundreds of rotations, and the wavelength dependence of starspot RV signals may not be consistent between stars due to differences in their magnetic fields and active regions.  We obtained precise multi-wavelength RVs of four rapidly-rotating M dwarfs (AD Leo, G 227-22, GJ 1245B, GJ 3959) using the near-infrared (NIR) Habitable-zone Planet Finder, and the optical Keck/HIRES spectrometer.  Our RVs are complemented by photometry from \kepp, TESS, and the Las Cumbres Observatory (LCO) network of telescopes.  We found that all four stars exhibit large spot-induced Doppler signals at their rotation periods, and investigated the longevity and optical-to-NIR chromaticity for these signals.  The phase curves remain coherent much longer than is typical for Sunlike stars.  Their chromaticity varies, and one star (GJ 3959) exhibits optical and NIR RV modulation consistent in both phase and amplitude.  In general, though, we find that the NIR amplitudes are lower than their optical counterparts.  We conclude that starspot modulation for rapidly-rotating M stars frequently remains coherent for hundreds of stellar rotations, and gives rise to Doppler signals that, due to this coherence, may be mistaken for exoplanets.

\end{abstract}
\keywords{techniques: radial-velocities, planets and satellites: fundamental parameters, stars: stellar activity}

\section{Introduction}
M dwarfs are attractive targets for radial velocity (RV) observations in search of exoplanets.  Their low masses ($M_* \leq 0.5 M_\odot$) result in larger Doppler amplitudes for a given planet mass, and the high density of absorption features in their spectra contain abundant RV information content.  The stellar-heated habitable zone \citep[SHZ;][]{kopparapu2013}, the range of orbital separations where the stellar insolation may allow surface water, is much closer to the star for M dwarfs.  Thus, SHZ planets orbiting M dwarfs have greater RV amplitudes, and may be discovered more quickly than for FGK stars.  The planet-to-star size ratio offers advantages for studying transiting planets as well; to date, the characterization of atmospheres for Earth-sized exoplanets has been largely restricted to planets orbiting late-type stars \citep[e.g.,][]{kreidberg2014}.

Doppler velocimetry of stars later than M3, however, is challenging for optical spectrographs.  Such stars are brightest and richest in RV information content at red-optical and near-infrared (NIR) wavelengths.  To this end, we have developed and commissioned the Habitable-zone Planet Finder, or HPF, an ultra-stable NIR Doppler spectrograph designed to provide sensitivity to low-mass exoplanets orbiting nearby M dwarfs \citep{mahadevan2012,mahadevan2014}.  HPF is a facility instrument on the 10\,m Hobby-Eberly Telescope at McDonald Observatory in Texas.  The combination of NIR wavelength coverage with a 10\,m aperture allows HPF to overcome the inherent faintness and redness of mid-late M stars and achieve suitable S/N for precision velocimetry.  

Designed from the bottom up to survey mid-to-late M-dwarfs at $1-3\unit{m/s}$ RV precision for planets in the SHZ, HPF covers the information-rich $z$, $Y$, and $J$ bands ($810 \unit{nm}$ to $1280 \unit{nm}$) at a resolving power of $R\sim 55,000$, where the M-dwarf SED peaks. For sensitivity in the near-infrared, HPF uses a Hawaii-2RG detector array with a 1.7$\unit{\mu m}$ cutoff, and its optical table is maintained at $180 \unit{K}$ to suppress the near-infrared background blackbody radiation.  To achieve its goal of $1 \unit{m/s}$ RV precision around bright M-dwarfs, HPF employs an active temperature control system capable of maintaining the HPF cryostat at $180 \unit{K}$ with $1 \unit{mK}$ stability long-term \citep{stefansson2016}.  HPF is wavelength calibrated by a custom NIR laser frequency comb (LFC) that provides calibration precision to better than $10 \unit{cm/s}$ precision \citep{metcalf2019}.

When searching for low-mass exoplanets with precise RVs, it is crucial to remain cognizant of astrophysical RV variability originating from time-dependent processes in the target stars' atmospheres.  Such variability is nearly always temporally correlated, and may be quasi-periodic, which can lead to false-positive exoplanet detections \citep[e.g.,][]{robertson2014}.  It is expected that observing in the NIR will reduce \citep{marchwinski2015}, but not eliminate \citep{reiners2010}, the impact of astrophysical variability due to the diminished contrast of stellar magnetic features (spots, plage, etc.) to the surrounding photosphere. An exoplanet candidate at its host star's rotation period has a high probability of being a false positive, as rotating stellar features such as starspots can create RV signals at the rotation period or its harmonics \citep{boisse2011,newton2016,vanderburg2016_prot}. This is especially true for young, active stars, where starspots may create RV signals with amplitudes in excess of $100 \unit{m/s}$ \citep[e.g.,][]{queloz2001}. 

\citet{tuomi2018} recently reported the detection of a candidate exoplanet with a period $P = 2.23$ days orbiting the M4.5V dwarf AD Leo, an active nearby ($d = 4.9 \unit{pc}$) star, based on archival RVs from the HARPS and HIRES spectrographs.  The candidate planet's period matches the rotation period of the star, suggesting that it is in spin-orbit resonance with its host star. The planet, if confirmed, would therefore be rare for a number of reasons. In addition to the unusual circumstance of being locked in spin-orbit resonance, it is more massive than is known to be typical of the M dwarfs surveyed to date \citep[e.g.,][]{endl2006}.  \citet{tuomi2018} detailed several arguments for why the RV signal of AD Leo is likely to be caused by an exoplanet, including the fact that the RV signal appears to remain coherent for the extent of the HIRES and HARPS observations, while more direct tracers of astrophysical variability such as photometry and activity-sensitive absorption lines show a highly incoherent rotation signal.  Nevertheless, out of an abundance of caution, Tuomi et al.~retained the ``candidate'' designation for the RV signal pending further investigation.  We observed AD Leo as one of the first HPF commissioning targets, under the hypothesis that if the 2.23-day RV signal is in fact caused by stellar activity, NIR RVs should readily reveal this via a reduction in RV amplitude. This technique has been employed at lower precision in the vetting of hot Jupiter candidates around very active T Tauri stars \citep{crockett2012,johnskrull2016}.

Our HPF observations of AD Leo and other active M dwarfs challenged the exoplanet hypothesis.  For several targets, we observed large-amplitude RV signals at the stellar rotation period that remained remarkably consistent in amplitude and phase over many stellar rotations.  For the case of the nearby M dwarf GJ 1245B, our earliest observations revealed a signal that was similar to periodic behavior in archival optical Keck/HIRES RVs.  Thus, it appears that young, active M dwarfs with persistent RV signals at the stellar rotation period may be common.

Long-lived RV signals at the rotation periods of M dwarfs are potentially consistent with either planets or stellar magnetic activity.  Coherent starspot signals from M stars have been observed to persist for years at a time \citep{robertson2014,davenport2015}, which could give rise to the observed RV signals.  On the other hand, there is at least one other candidate giant planet orbiting a young M star very close to the stellar rotation period \citep[PTFO 8-8695;][]{vaneyken2012,koen2015,ciardi2015b,yu2015,tanimoto2020}.  However, the fact that we have identified several such signals in the first few HPF targets would imply that spin-orbit coupled giant planets are common around M dwarfs, which is inconsistent with exoplanet surveys of older, quiet M stars. Exoplanet occurrence rates from transits \citep{dressing2015} and RVs \citep{endl2006} conclusively rule out a large population of close-in giant planets for M dwarfs.  Most recently, \citet{hsu2020} used \kep statistics to place an 84\% upper limit of 3-4\% for the fraction of late-M stars hosting planets with $R > 4R_{\oplus}$ at periods less than 4 days.  Thus, if spin-orbit coupled gas giants are common for young M stars, they must somehow be destroyed as the star ages in order to have avoided detection by previous surveys.

Coincidentally, we noticed that AD Leo and several of our other early HPF targets are similar in that their rotational velocities (\vsinip) and rotation periods imply they are viewed at low inclinations, sometimes close to pole-on.  This caused us to speculate that the observed RV signals might be due to a type of long-lived feature in the stellar atmospheres that preferentially occurs at high latitudes.

In an attempt to determine the origin of the persistent RV signals for AD Leo and other pole-on M dwarfs, we obtained precise optical and NIR RVs of 4 such stars using Keck/HIRES and HET/HPF, respectively.

As we were preparing this manuscript, \citet{carleo2020} published multi-wavelength RVs of AD Leo from GIANO-B+HARPS-N that ruled out the exoplanet origin for the 2.23-day Doppler signal.  Our results are complementary to theirs, as we will discuss in further detail in the relevant sections.


\begin{table*}
\begin{center}
\footnotesize
\begin{tabular}{l | c c c c }

\hline
Parameter & AD Leo & G 227-22 & GJ 3959 & GJ 1245B \\
\hline
&\multicolumn{4}{ c }{} \\
\emph{Measured Quantities} & \multicolumn{4}{ c }{} \\
&\multicolumn{4}{ c }{} \\
Parallax $\pi~(\unit{mas})^{(1)}$ & $201.3683 \pm 0.0679$ & $128.4871 \pm 0.0576$ & $88.8598 \pm 0.0771$ & $214.5285 \pm 0.0824$ \\
Apparent $G_{BP}$ magnitude$^{(1)}$ & $9.628 \pm 0.006$ & $13.723 \pm 0.006$ & $15.118 \pm 0.009$ & $14.353 \pm 0.006$ \\
Apparent $G_{RP}$ magnitude$^{(1)}$ & $7.053 \pm 0.005$ & $10.459 \pm 0.005$ & $11.589 \pm 0.005$ & $10.511 \pm 0.005$ \\
Apparent $K$ magnitude$^{(2)}$ & $4.593 \pm 0.017$ & $7.652 \pm 0.02$ & $8.506 \pm 0.016$ & $7.387 \pm 0.018$ \\
Rotational velocity \vsini (km/s) & $2.63 \pm 0.6^{(3)}$ & $11.3 \pm 1.5^{(4)}$ & $7.1 \pm 1.5^{(4)}$ & $6.8 \pm 1.9^{(5)}$ \\
Rotation period $P_{rot}$ (d) & $2.2399 \pm 0.0006^{(6)}$ & $0.28018 \pm 1 \times 10^{-5}$ $^{(7)}$ & $0.51207 \pm 5 \times 10^{-5}$ $^{(7,*)}$ & $0.709 \pm 0.001^{(8)}$ \\
&\multicolumn{4}{ c }{} \\
\emph{Derived Quantities} & \multicolumn{4}{ c }{} \\
&\multicolumn{4}{ c }{} \\
Distance $d~(\unit{pc})^{(9)}$ & $4.965 \pm 0.002$ & $7.781 \pm 0.004$ & $11.25 \pm 0.01$ & $4.661 \pm 0.002$ \\
Absolute $K$ magnitude $M_K$ & $6.11 \pm 0.02$ & $8.20 \pm 0.02$ & $8.25 \pm 0.02$ & $9.04 \pm 0.02$ \\
Mass $M_*~(M_\odot)^{(10)}$ & $0.43 \pm 0.02$ & $0.16 \pm 0.01$ & $0.16 \pm 0.01$ & $0.11 \pm 0.01$ \\
Radius $R_*~(R_\odot)^{(11)}$ & $0.428 \pm 0.003$ & $0.195 \pm 0.002$ & $0.191 \pm 0.001$ & $0.142 \pm 0.001$ \\
Effective Temperature $T_{\mathrm{eff}}~(\unit{K})$ & $3382 \pm 80^{(3,11)}$ & $3124 \pm 51^{(12)}$ & $3008 \pm 66^{(13)}$ & $2859 \pm 60^{(11)}$ \\
Luminosity $L_*~(L_\odot)$ & $0.0215 \pm 0.002$ & $0.0033 \pm 0.0003$ & $0.0027 \pm 0.0003$ & $0.0012 \pm 0.0001$ \\
Stellar inclination $i$ ($^\circ$)$^{(14)}$ & $17 \pm 4$ & $19 \pm 3$ & $23 \pm 5$ & $51_{-15}^{+20}$ \\
Conservative HZ Bounds (AU)$^{(15)}$ & 0.15---0.3 & 0.06---0.12 & 0.05---0.11 & 0.04---0.07 \\
Isochrone Age (Myr)$^{(16)}$ & $28^{+7}_{-5}$ & $500^{+1100}_{-340}$ & $150^{+90}_{-50}$ & $290^{+60}_{-50}$ \\

\hline
\end{tabular}
\caption{\footnotesize Measured and derived stellar properties for our targets.  References: (1): \citet{gaiaDR2}, (2): \citet{cutri2003}, (3): \citet{houdebine2016}, (4): \citet{reiners2018}, (5): \citet{delfosse1998}, (6): \citet{morin2008}, (7): This work, (8): \citet{lurie2015}, (9): \citet{bailerjones2018}, (10): \citet{delfosse2000}, (11): \citet{mann2015}, (12): \citet{schweitzer2019}, (13): \citet{muirhead2018},      (14): \citet{masuda2020}, (15): \citet{kopparapu2013}, (16): \citet{morton2015}, (17): \citet{engle2018}. \\
*Derived from RVs}
\label{tab:stellar}
\end{center}
\end{table*}

\section{Target Selection}
\label{sec:targets}

In order to investigate whether pole-on M stars are particularly likely to exhibit long-lived RV signals at the stellar rotation period, we assembled a list of ``AD Leo analogs" in the northern hemisphere.  Specifically, we sought to identify rapidly-rotating mid-to-late M dwarfs with low inclination angles ($i$).

We adopted preliminary inclination values based on literature estimates of stellar \vsini values and rotation periods. We estimated the stellar inclinations using the methodology in \cite{masuda2020}, which correctly accounts for the correlation in the stellar equatorial velocity ($v_{\mathrm{eq}}$)---estimated from the rotation period $P_{\mathrm{rot}}$ and the stellar radius $R_*$ as $v_{\mathrm{eq}} = 2\pi R_* / P_{\mathrm{rot}}$---and its projection on the sky ($v \sin i$). We selected rapid rotators with $i \lessapprox 20^{\circ}$ that could be observed from HET and Keck. In this section, we provide a brief description of the 4 targets we selected. Relevant stellar parameters of these targets are presented in Table \ref{tab:stellar}. Updated parallax measurements from GAIA DR2 \citep{gaiaDR2} facilitate refinements of our targets' fundamental stellar parameters, particularly those that rely on empirical calibrations to the absolute $K$-band magnitude, $M_K$. Specifically, we have used the $M_K$ calibrations of \citet{delfosse2000} and \citet{mann2015} to obtain estimates of the stellar mass and radius, respectively, in the Table. In cases where the analysis presented herein has yielded a more precise stellar rotation period than previously published, we have adopted that value in Table \ref{tab:stellar} and updated the stellar inclination accordingly.

The rapid rotation of our targets suggests they are likely young, although estimating precise ages for M stars is notoriously difficult due to the slow evolution of their fundamental parameters \citep[e.g.][]{laughlin1997} and the apparent rapid transition from fast to slow rotation states \citep{newton2017}.  We have estimated ages for our targets using the stellar isochrone fitting package \texttt{Isochrones} \citep{morton2015}.  We used Gaia DR2 parallaxes and $G_{BP/RP}$ colors, along with $K$-band magnitudes and $T_{eff}$ estimates from Table \ref{tab:stellar} as inputs.  We used \texttt{Isochrones}' default priors, except for the prior on stellar age, for which we adopted a uniform prior on $\log$(age) from 6 to 10.  The resulting ages and their associated uncertainties are listed in Table \ref{tab:stellar}.  While it is especially difficult to compare rotation-based age estimates for stars outside a physically-associated collection such as a moving group, our isochrone-based ages are consistent with the rotation-age relation of \citet{engle2018}.  Broadly speaking, all indications are that our targets are all less than 1 Gyr old.

In Table \ref{tab:stellar}, the uncertainties on $\pi$, $K$, \vsinip, $P_{rot}$, $d$, and $T_{\mathrm{eff}}$ are shown as quoted in the references listed.  The uncertainties on rotation periods measured herein are described in \S\ref{sec:analysis}, and the uncertainties on the stellar inclination $i$ indicate the 68\%-credible interval resulting from the \citet{masuda2020} technique. For all other quantities, the 1$\sigma$ uncertainties are taken from basic propagation of the measurement errors on the parameters used to infer those values.

\subsection{AD Leo}
\label{sec:adleo}
At just under 5 parsecs from Earth, AD Leo is one of the closest and most well-studied M dwarfs.  Its high levels of activity and evolving magnetic field have been well studied, including its flares \citep[e.g.,][]{hawley2003,vdb2003}, starspots \citep{huntwalker2012}, and global magnetic field \citep{morin2008,lavail2018}.

AD Leo's mass of $0.43 M_\odot$ places it comfortably in the regime of partially convective stars according to the models of \citet{chabrier1997}.  Thus, its magnetic field is likely produced by the $\alpha\Omega$ dynamo that results from shearing at the boundary between its convective exterior and radiative interior \citep{thompson2003}.  The $\alpha\Omega$ dynamo is believed to power long-period magnetic cycles in partially-convective stars \citep{brown2011}, and indeed \citet{buccino2014} and \citet{tuomi2018} observe a 7-year activity cycle in ASAS photometry of AD Leo.  Interestingly, \citet{morin2008} find that AD Leo exhibits significantly lower magnetic \emph{flux} than stars that have similar magnetic \emph{fields} but which have masses below the $M_* = 0.35 M_\odot$ boundary for fully-convective objects.  They attribute this behavior to the more efficient generation of a global magnetic field in AD Leo, which is again consistent with the presence of an $\alpha\Omega$ dynamo.

\subsection{G 227-22}
G 227-22 is a fully-convective M dwarf near the Northern Ecliptic Pole, and as such receives multi-sector coverage from TESS.  The long-baseline TESS coverage presented a unique opportunity to study the effects of stellar magnetic activity on both photometry and spectroscopy, so we observed this star in the early stages of HPF science operations.  Our HPF RVs revealed a remarkably consistent signal at the star's 0.28-day rotation period \citep{newton2016}.  Given that AD Leo and G 227-22 are both active, nearly pole-on M dwarfs, we considered whether their persistent RV signals might have similar astrophysical origins.  The similarities between these two stars prompted the multi-waveband study presented here.

\subsection{GJ 3959}
GJ 3959 is the faintest and most distant of our selected targets.  As such, it is less well studied than the rest of our sample.  Its fundamental properties are very similar to those of G 227-22, but its rotation period is a factor of 2 longer.  Its $\sim 0.5$-day rotation period is an upper limit for achieving useful phase coverage given our Keck/HIRES observing strategy of attempting to cover a full rotation over a single night.

\subsection{GJ 1245B}
The M6 dwarf GJ 1245B is the smallest and coolest of our targets.  It is the 43rd-closest star to the Sun \citep{gaiaDR2}, and a member of the GJ 1245 hierarchical triple system of M dwarfs.  The system consists of the GJ 1245AB binary, orbited by the faint M8 companion GJ 1245C \citep{harrington1990}.

\citet{lurie2015} analyzed the \kep lightcurve of the GJ 1245AB binary, isolating periodic signals associated with the rotation of each component.  They determined GJ 1245B has a rotation period of 0.709 days.  Combined with a \vsini of 6.8 km/s \citep{delfosse1998}, this implies a stellar inclination of $51^\circ$, which is considerably higher than our desired maximum of $i \sim 20^\circ$.  However, the availability of archival HIRES RVs---which showed evidence of a signal at the stellar rotation period---made this star an opportunistic target, for which we essentially only needed to acquire new HPF velocities.

\section{Observations and Data Reduction}
\label{sec:data}
Our targets all have significant amounts of spectroscopic and photometric data available.  We obtained some of these data specifically for this experiment, but also took advantage of archival public data.  Our results rely primarily on the following sources of data.

\subsection{HPF radial velocities}
We obtained NIR radial velocities of our targets using HPF in its standard queue-scheduled mode. We used typical total exposure times of about 15 minutes, usually divided into 3 5-minute exposures.  

The wavelength solutions for the HPF spectra are calibrated using the laser frequency comb \citep[LFC;][]{metcalf2019}. HPF's calibration fiber offers the ability to obtain an LFC spectrum simultaneously with our science exposures, but we only used this option for AD Leo and G 227-22. For the other two targets, GJ 3959 and GJ 1245B, we corrected for any wavelength drift using the LFC exposure frames taken throughout the night following the methodology described in \cite{stefansson2020}. This method has been shown to result in drift correction errors at the sub-m/s level, which is much smaller than the RV amplitudes studied here.

Our HPF observations are spread throughout the 2018 and 2019 observing seasons---more details on the sampling of each target are provided in \S\ref{sec:analysis}. We were allocated time to obtain one HPF observation each of GJ 3959 and G 227-22 on the same nights as our Keck observations in order to anchor the RV zero points in any models of the observed signals as exoplanet orbits.  However, weather prevented this observation in the case of G 227-22.

The HPF 1D spectra were reduced and extracted with the custom HPF data-extraction pipeline following the procedures outlined in \cite{ninan2018}, \cite{kaplan2018}, and \cite{metcalf2019}. To compute the radial velocities from HPF spectra, we adapted the publicly available \textit{SpEctrum Radial Velocity Analyzer} code \cite{zechmeister2018} to reduce the HPF 1-dimensional spectra following the methodology in \citep{stefansson2020}. For the RV calculations, we mask out both telluric lines, and sky emission lines. 

\subsection{HIRES radial velocities}
Our NIR RVs from HPF are complemented with optical RVs, primarily from the HIRES spectrometer on the 10\,m Keck I telescope.  For AD Leo and GJ 1245B, previously-available public data were sufficient, while we obtained new RV data for GJ 3959 and G 227-22.

\subsubsection{Archival HIRES RVs}
Time-series HIRES spectra of AD Leo and GJ 1245B are available on the Keck Observatory Archive (KOA).  RVs for these spectra were computed and provided by \citet{butler2017}.  We have used the RV corrections provided by \citet{talor2019}, which remove small systematic errors present in the Butler et al. velocities.  All RV signals analyzed herein are large enough to be insensitive to these small corrections, but we nevertheless used the most up-to-date reduction.  We have excluded from our analysis any observations for which the median photon counts per pixel fell below 650.

\subsubsection{New HIRES RVs}
We obtained new HIRES RVs of G 227-22 and GJ 3959 as part of this experiment.  Because Keck is not queue-scheduled, we chose these two targets in part because their rotation periods are short enough that we could cover most or all of a full stellar rotation in a single night.

Our observations took place on the nights of 2019 May 17 and 2019 June 8 (UT).  Our observing strategy consisted of observing each star at high cadence over the course of a night.  For GJ 3959, we obtained 11 RVs on the night of May 17, and 4 RVs on the night of June 8.  We acquired 40 RVs of G 227-22, all on the night of June 8.

We configured HIRES in the ``red" cross-disperser, which provides useful wavelength coverage from approximately $3600-8000$ \AA.  We used a slit width of $0.861 \arcsec$, yielding a resolving power $R \sim 50000$ near $\lambda = 5000$ \AA.  Precise RVs are obtained by placing a temperature-controlled cell of molecular iodine (I$_2$) vapor in front of the slit.  The I$_2$ imprints a series of weak absorption lines over the stellar spectrum, which can be used to precisely calibrate the wavelength solution and track changes in the instrument profile that would otherwise cause shifts in our measured RVs \citep{valenti1995,butler1996}.  We performed basic data reduction (bias subtraction, flat fielding, etc.) and spectral extraction using custom IRAF scripts, and extracted precise RVs using the AUSTRAL \citep{endl2000} pipeline.

\subsection{HARPS radial velocities}
Our analysis includes RVs of AD Leo from the HARPS spectrograph on the 3.6\,m Telescope at La Silla.  We adopt the HARPS RVs as presented in \citet{trifonov2020}, which have been reduced using the SERVAL \citep{zechmeister2018} pipeline, and corrected for small night-to-night systematics as done by \citet{talor2019} for the archival HIRES RVs.  

\subsection{LCO photometry}

To better understand and analyze astrophysical RV variability, the HPF survey is also conducting a Key Project titled ``High-Cadence Monitoring of the Sun's Coolest Neighbors'' on the Las Cumbres Observatory \citep[LCO;][]{brown2013} global telescope network (previously known as LCOGT).  The project primarily uses LCO's network of 0.4\,m telescopes to obtain multi-color photometric observations of M stars targeted by HPF at a high cadence--typically once per 24 hours.  Our goal is to identify signals associated with stellar activity that may propagate to RV measurements, and to evaluate the color dependence of these features.  We observe our targets in the Johnson $V$ and SDSS $i^\prime$ filters in order to monitor the stellar variability in the optical and NIR bandpasses.  Our cadence scheduling includes an airmass limit of 2.0 to minimize scintillation errors, and systematics due to differential atmospheric extinction.  We also do not observe a target if it is within $30^\circ$ of the Moon.  

Basic image reduction (e.g., flat, dark correction) for LCOGT images is performed automatically with the observatory's \textit{Bansai} pipeline. Using these reduced images, we performed differential photometry with the AstroImageJ \citep{collins2017} analysis package.  We used AstroImageJ's variable aperture setting, in which the program calculates the point-spread function of each image and scales the photometric aperture accordingly.  This adjustment accounts for variable seeing, focus position, and tracking errors from one image to another.  We manually removed any clear outliers ($>5\sigma$), which were clearly due to cosmic rays or insufficient signal-to-noise.

A number of complications are associated with using an automated, multi-site facility such as LCO.  We find that the detectors on the 0.4\,m network exhibit intermittently hot pixels, which naturally vary from site to site.  The CCD noise pattern also changed in June 2018, when the observatory switched from $2\times2$ pixel binning to $1\times1$.  The pointing and tracking of the 0.4\,m telescopes are not perfectly consistent, so the target and comparison stars may fall on unreliable pixels in some images and not others.  The magnitude of the pointing shifts is large enough that some of our comparison stars occasionally fall off the field of view, further complicating our data reduction.  We handle these correlated noise sources using the Inhomogeneous Ensemble Photometry (IEP) technique \citep{honeycutt1992}, which uses a least-squares solution to remove non-astrophysical variability across a series of images.  IEP has the advantage of not requiring a given comparison star to appear in every image.  After applying the IEP algorithm, we fit and subtracted zero-point offsets to values taken with a given telescope and binning ($2\times2$ versus $1\times1$) mode.

\subsection{\kep photometry}
GJ 1245B lies within the \kep field, and was observed as part of \kepp's 4-year primary mission to identify transiting exoplanets \citep[e.g.,][]{borucki2010}.  The GJ 1245 binary is unresolved in \kep images, which complicates the time-series analysis of the lightcurve.  \citet{lurie2015} analyzed the \kep photometry of the GJ 1245 system, finding a 0.71-day rotation period for GJ 1245B.  We have not reanalyzed the \kep data, and instead rely on the Lurie et al.~analysis herein.

\subsection{TESS photometry}
As of this writing, G 227-22 and GJ 1245B have been observed by the TESS all-sky photometric survey \citep{ricker2015} in the 2-minute ``short" cadence mode.  GJ 1245B was observed in TESS Sectors 14 and 15, for a total observational time baseline of 54 days.  G 227-22 is scheduled to be observed in all of Sectors 14-25, and we have currently analyzed data from Sectors 14-20.  GJ 3959 will be observed in Sectors 24-25 (April-June 2020), while AD Leo is not scheduled to be observed in the TESS prime mission.  

We use the standard Presearch Data Conditioning (PDC or PDCSAP) flux values as provided by the TESS pipeline.  We used \lightkurve \citep{lightkurve2018} to retrieve the TESS lightcurves, as well as to perform outlier rejection and binning.


\section{Analysis}
\label{sec:analysis}

\subsection{AD Leo}

\begin{figure}
    \centering
    \includegraphics[width=\columnwidth]{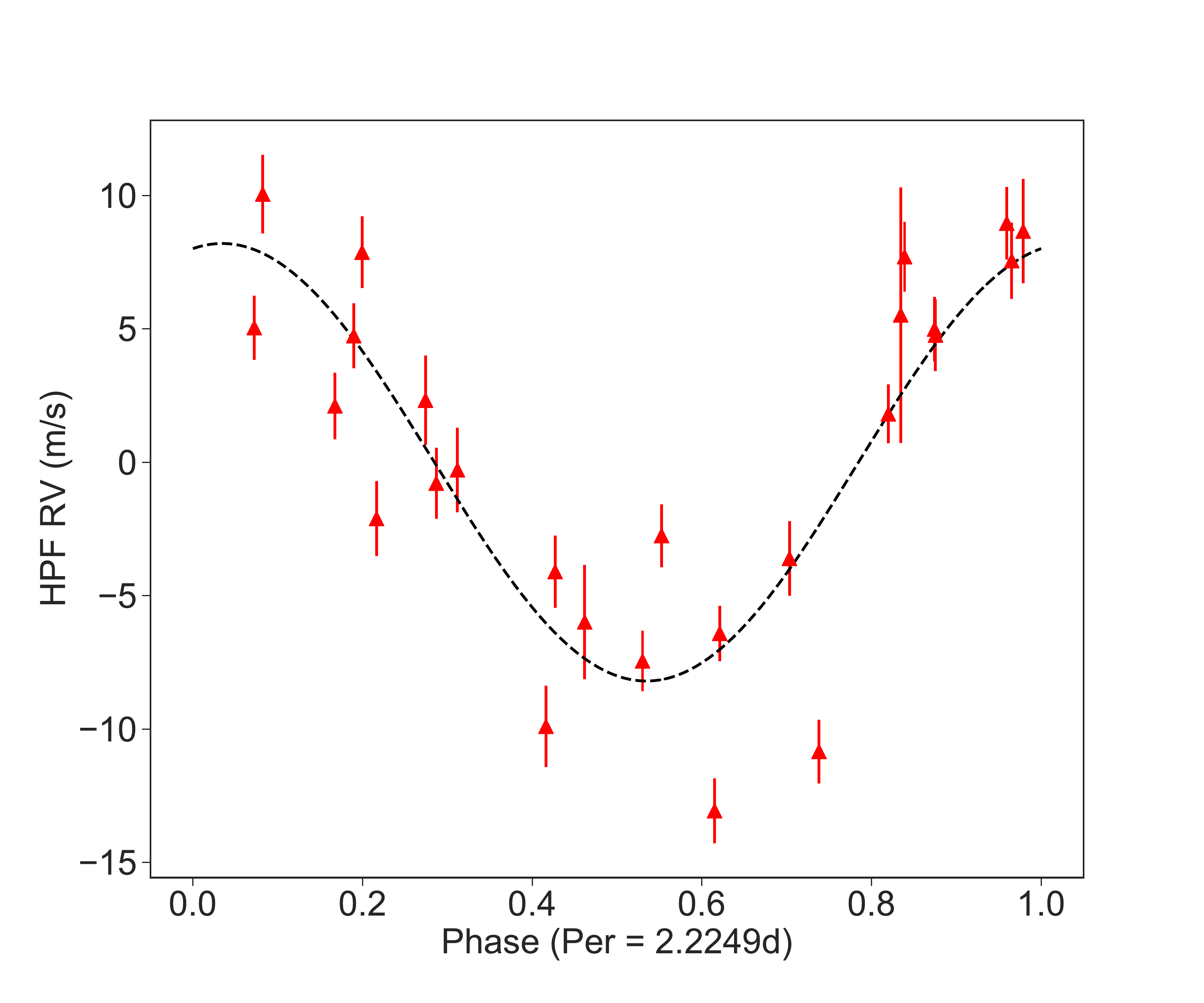}
    \caption{HPF RVs of AD Leo from the 2019 observing season, folded to the best-fit rotation signal (dashed line).  The 8.2 \unit{m/s} amplitude is less than half the proposed amplitude of the planet proposed by \citet{tuomi2018}.}
    \label{fig:gj388_hpf}
\end{figure}

\begin{figure*}
\centering
\includegraphics[width=\textwidth]{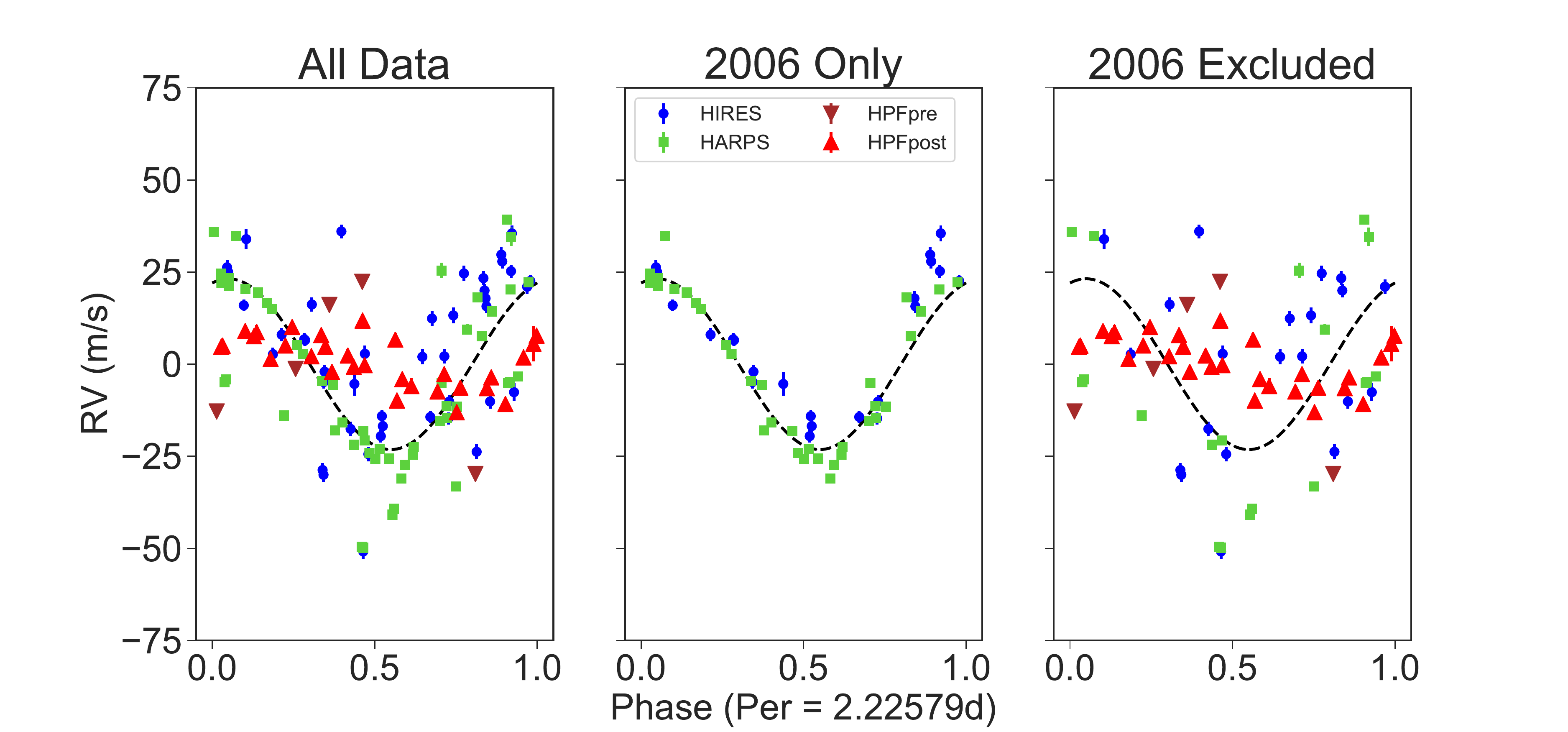}
\caption{RVs of AD Leo, folded to the proposed planet period from \citet{tuomi2018}.  The left panel shows all RVs, while the other panels show the RVs from the 2006 season (middle), and RVs acquired in seasons other than 2006 (right).  In each panel, the Tuomi et~al.~model is shown as a dashed line.  Only the 2006 data are consistent with the Tuomi et~al.~model.  The 2019 HPF RVs (HPFpost) in particular are incompatible in both amplitude and phase with the proposed exoplanet signal.}
\label{fig:gj388_2006}
\end{figure*}

\subsubsection{Stellar rotation}
AD Leo's rotation period of 2.23 days is well established by a number of spectroscopic and photometric analyses \citep[e.g.,][]{morin2008,huntwalker2012}.  This rotation period, combined with estimates that the star's rotational velocity was approximately $v \sin i \sim 3$ km/s \citep{morin2008,houdebine2016}, suggested the star was viewed close to pole-on. 

\subsubsection{New RVs}
We have obtained 35 HPF RVs between April 2018 and February 2020. Our 2018 RVs were obtained during HPF commissioning, after which the instrument's vacuum chamber was opened for maintenance before the start of full science operations. We observed an RV zero-point offset in our data after this opening, and therefore treat our RVs before (HPFpre) and after (HPFpost) as separate data sets.  After fitting an offset between the HPFpre and HPFpost RVs, we find an RMS scatter of 9.5 \unit{m/s}, with a mean single-measurement uncertainty of 1.5 \unit{m/s}. The raw HPF variability is in contrast with the HARPS+HIRES time series; after fitting and removing an offset between the HARPS and HIRES RVs, we find an RMS of 22 \unit{m/s}. The difference in variability cannot be attributed to a lack of precision in the optical spectrometers: the HARPS RVs have a mean uncertainty of 1.1 \unit{m/s}, while the HIRES mean uncertainty is 1.9 \unit{m/s}.  These uncertainties are unlikely to be significantly underestimated, as the $\sim1$ \unit{m/s} precision achieved by these instruments (and pipelines) on extremely quiet mid-M dwarfs such as GJ 581 \citep{vogt2010} and Barnard's Star \citep{ribas2018} demonstrate their performance on very cool stars.

In addition, the HARPS RVs of AD Leo provided by \citet{trifonov2020} include 5 velocities from April 2016 that were not considered in the \citet{tuomi2018} analysis.  Our results do not depend on whether we use these data, but we include them for the sake of completeness.

\subsubsection{Evaluating the exoplanet hypothesis}
A key advantage of HPF's NIR wavelength coverage is that it offers a way to determine whether a periodic RV signal is caused by an exoplanet or spot modulation via the ratio of optical-to-NIR RV amplitudes.  The lower spot-photosphere contrast in the NIR should lead to a smaller RV amplitude for a starspot signal relative to the optical \citep{reiners2010,marchwinski2015}, while true Keplerian motion should be achromatic.

Upon comparing the optical and NIR RVs for AD Leo, it is clear that the star does not exhibit a large-amplitude, achromatic signal at the rotation period.  The RV signal at the rotation period is inconsistent between the optical and NIR RVs, and even between seasons for the HPF RVs---the HPF RVs exhibit a much higher RMS scatter in the 2018 season (23 \unit{m/s}) than in 2019 (6.4 \unit{m/s}).  This drop in variability from 2018 to 2019 is also observed by \citet{carleo2020} in visible-band HARPS-N RVs.  In Figure \ref{fig:gj388_hpf}, we show the 2019 HPF RVs, folded to the best-fit model at the rotation period \citep[as identified by \radvelp,][]{fulton2018}.  While the rotation signal is present in the data, its amplitude and phase are broadly inconsistent with the planet proposed by \citet{tuomi2018}; the amplitude is only 8.2 \unit{m/s}, as opposed to the $\sim 20$ \unit{m/s} amplitude modeled by Tuomi et~al.

The seasonal variability of the rotation signal in the RVs is inconsistent with the claim by \citet{tuomi2018} that the signal remained coherent over more than 10 years of HIRES and HARPS observations.  However, that claim was supported by comparing subsets of HARPS observations taken over a span of less than 3 months, whereas we find for our targets that starspot signals can remain coherent for much longer.  When looking at the entire optical time series, we find that the rotation signal does not appear to be truly coherent.  In particular, more than half of all the optical RVs were taken within the 2006 observing season.  In Figure \ref{fig:gj388_2006}, we show all available RVs phased to the planet period proposed by \citet{tuomi2018}.  We have then separated the data into RVs from the 2006 season, and from all other seasons.  The non-2006 data is clearly inconsistent with the Tuomi et~al.~model; they exhibit an RMS scatter of 20.2 \unit{m/s} around the 2.22579-day period, as opposed to 5.8 \unit{m/s} in 2006.  Furthermore, there is no frequency near the stellar rotation period for which all RVs can be modeled with a coherent signal.  This is most clearly evidenced by the amplitude change between the 2006 optical and 2019 NIR signals.  Thus, based on the incompatibility of the NIR HPF RVs with the Tuomi et~al.~model, and the season-to-season incoherence of all RVs, we conclude that the 2.23-day signal seen in RVs of AD Leo is caused by starspot activity rather than a spin-orbit-coupled exoplanet.

\subsection{GJ 1245B}

\begin{figure*}
    \centering
    \includegraphics[width=\textwidth]{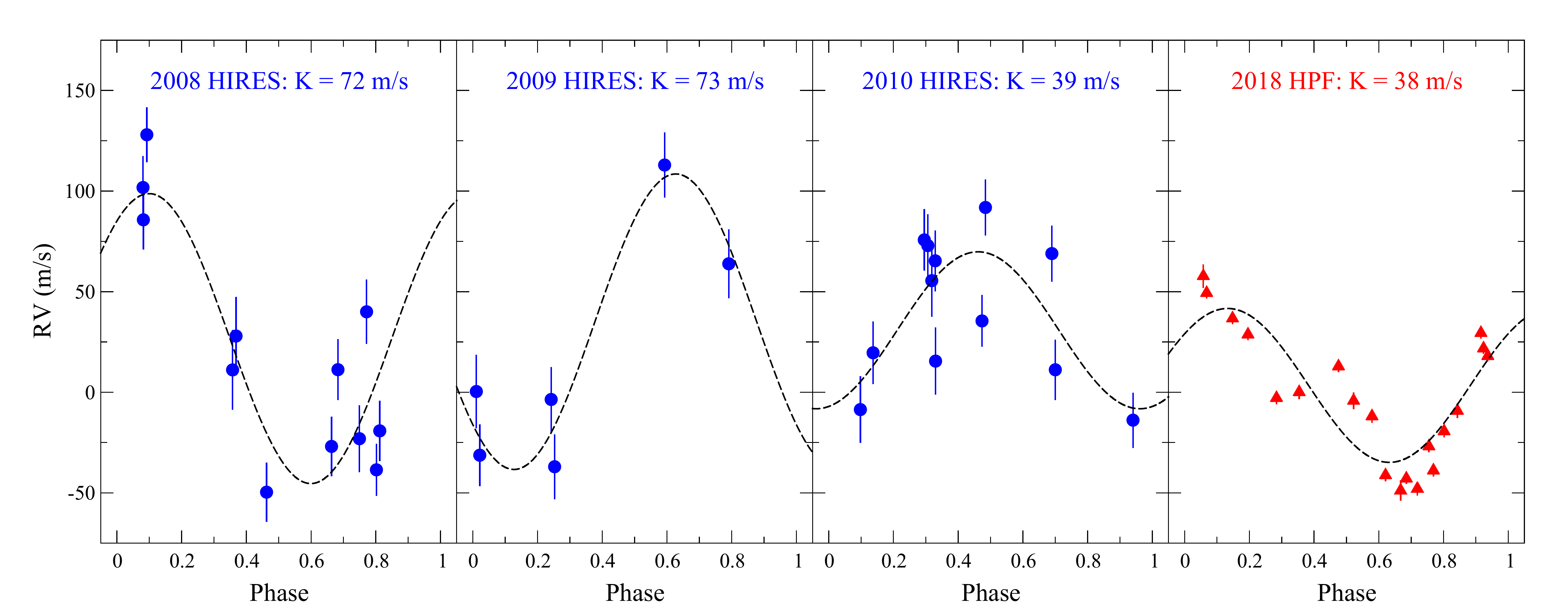}
    \caption{RVs of GJ 1245B, separated by season and phased to the 0.709-day rotation period identified by \citet{lurie2015}.  The rotation signal is typically coherent across a single observing season, but evolves in phase and amplitude from season to season.}
    \label{fig:gj1245B_seasons}
\end{figure*}

\subsubsection{Stellar rotation}

As mentioned previously, the \kep lightcurve of GJ 1245B has already been extensively analyzed in \citet{lurie2015}.  That study found that the star's photometric variability could be explained by three distinct, long-lived spots or spot complexes, and that the phase of the photometric variability sometimes shifted in phase over the span of 6 months to 1 year.  

\subsubsection{Available RVs}
\citet{talor2019} provide 41 RVs of GJ 1245B spanning 2005-2011.  These velocities have an RMS of 55.6 \unit{m/s} with a mean error of 17 \unit{m/s}.  The HIRES RVs from 2009 and 2010 are contemporaneous with the \kep photometry analyzed in \citet{lurie2015}.

Our 21 new HPF RVs of GJ 1245B cover approximately 2 months between mid-September and mid-November of 2018, with a single observation taken in May 2019. As noted for AD Leo, the NIR HPF RVs have a significantly lower RMS scatter than the optical RVs at 35.1 \unit{m/s} (mean error = 4.5 \unit{m/s}).

\subsubsection{RV analysis}
The RVs are clearly modulated by the 0.71-day stellar rotation.  When folding the HPF velocities to the 0.709-day period identified by \citet{lurie2015}, we find that they can be modeled with a single sinusoid, but that the 2019 observation is by far the most discrepant from the model, falling 37 \unit{m/s} ($\sim 7\sigma$) below the model expectation.  This suggests that the rotation signal has changed amplitude and/or phase between the two observing seasons.

Seasonal variability of the rotation signal is clear when examining the archival HIRES RVs alongside those from HPF. The combined RV time series is inconsistent with a coherent sinusoid at the rotation period, but RVs from individual seasons are well-described by a coherent signal. In Figure~\ref{fig:gj1245B_seasons}, we show the RVs separated by season, folded to the 0.709-day rotation period.  We have omitted individual points from 2005, 2006, 2011, and 2019 due to those seasons having too few observations.  For each season, the period has been fixed to the value provided by \citet{lurie2015}, but we have modeled a sinusoid with the amplitude, phase, and zero point as free parameters.  The amplitude and phase of the signal changes significantly over the 14-year combined time baseline, but the signal is consistently coherent over a single season, where a typical observing season covers 3-6 months.  This result is fully consistent with the 6-12-month phase evolution observed in the \kep lightcurve by \citet{lurie2015}. In fact, Lurie et.~al.~observe a significant change in the \kep phase curve of GJ 1245B at BJD$\sim2455233$ (see their Figure 7), which falls directly between the 2009 and 2010 observing seasons for HIRES. 

The RV amplitude of the rotation signal varies from 38 to 72 \unit{m/s}, with the HPF RVs having the lowest amplitude.  However, the amplitude of the 2018 HPF RVs is statistically indistinguishable from that of the 2010 HIRES velocities.  The 2010 RVs sample a relatively short-lived spot configuration; the \kep lightcurve shows the 2010 spot configuration decayed over about 1 year, and thus has no bearing on the 2018 RVs.  Additionally, the 2010 phase curve showed the lowest amplitude in the \kep lightcurve from \citet{lurie2015}, explaining the reduced RV amplitude.  Given that the 2010 observations clearly sample a period of reduced stellar variability, we speculate that the NIR amplitude of starspot-modulated RV signals for GJ 1245B is smaller than the optical amplitude for a given spot size, and that the spot configuration from 2018 is likely more similar to those observed in 2008 and 2009.

\subsection{G 227-22}
\label{sec:g227-22}

\begin{figure*}
    \centering
    \includegraphics[width=\textwidth, trim=0 8cm 0 8cm]{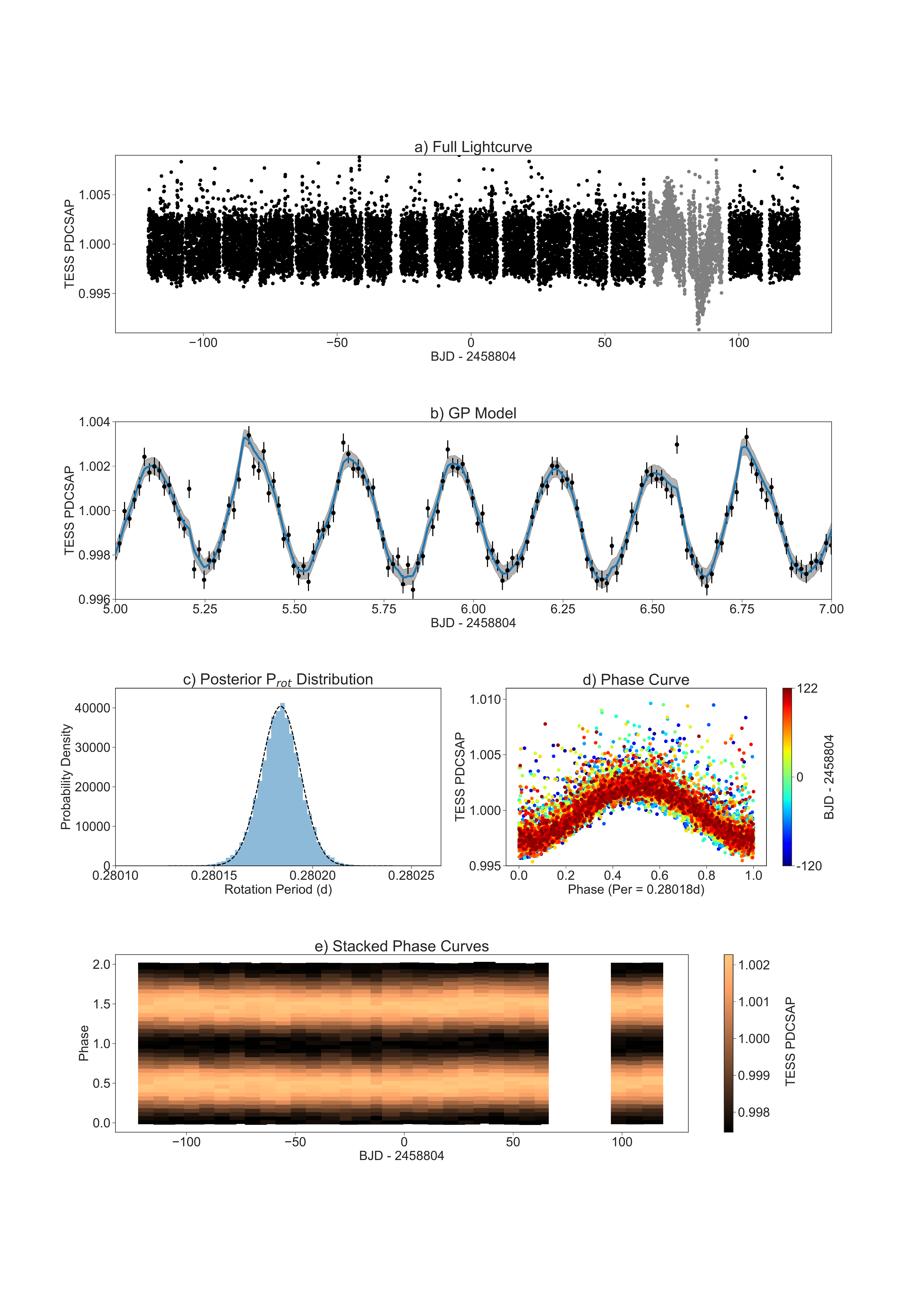}
    \caption{\footnotesize Analysis of rotationally-modulated TESS photometry of G 227-22 from Sectors 14-22.  The full lightcurve, binned and sigma-clipped, is shown in \emph{a}.  Data from Sector 21 are shown in gray to indicate we have excluded it from our analysis.  In \emph{b}, we show our GP model to the lightcurve for a representative segment of the time series.  The GP model is shown as a blue line, with the $1\sigma$ uncertainty region in gray.  \emph{c} shows a histogram of the 1D posterior distribution for the rotation period, and a best-fit Gaussian model.  \emph{d} shows the lightcurve folded to the 0.28018-day rotation period.  Points are color-coded according to their timestamps to show that the signal shows no significant evolution over TESS's time baseline.  In \emph{e}, we show stacked phase curves, median-binned into segments of 7 days and phase steps of 0.05.  Again, we see no phase drift in the lightcurve.}
    \label{fig:g227-22_rotation}
\end{figure*}

\subsubsection{Determining the rotation period}
The TESS lightcurve for G 227-22 exhibits frequent strong flare activity, as expected for such a rapidly rotating star.  Upon excluding the largest flares using iterative sigma clipping in \lightkurvep, the 0.28-day rotation period reported in \citet{newton2016} appears prominently in a power spectrum of the lightcurve.  As shown in Figure~\ref{fig:g227-22_rotation}, folding the lightcurve to the rotation period reveals the signal is coherent throughout TESS Sectors 14-22, corresponding to about 900 stellar rotations.  The rotation signal exhibits a photometric amplitude of approximately 0.23\%.

As shown in Figure \ref{fig:g227-22_rotation}a, the TESS photometry from Sector 21 contains a significant amount of correlated noise that is not present in data from other sectors.  We suspect it may be due to some source of scattered light, as observed by \citet{dalba2020}.  While the rotation signal from Sector 21 is consistent with the other sectors, we have omitted the Sector 21 photometry from our analysis in order to avoid additional uncertainty due to the unique systematics during that time.

The regularity of the photometric phase curve allows us to very precisely determine the stellar rotation period.  As suggested by \citet{angus2018}, we estimated the rotation period by modeling the TESS lightcurve using a Gaussian process \citep[GP;][]{ambikasaran15} correlated noise model, which uses a kernel function $k_{ij} = k(t_i - t_j)$ to constrain the covariance between two given points in a time series.  As is becoming typical when modeling stellar RV or photometric variability, Angus et al.~determined rotation periods from \kep lightcurves using a quasi-periodic kernel.  For computational efficiency, we used the kernel designed by \citet{dfm2017} as a drop-in replacement for use with the \celerite scalable 1D GP formulation:

\begin{equation*}
\begin{aligned}
    k_{ij} = & \frac{B}{2 + C} \exp{-\bigg(\frac{|t_i-t_j|}{L}\bigg)} \bigg[\cos{\bigg(\frac{2\pi|t_i-t_j|}{P_{\textrm{rot}}}\bigg)} + (1 + C)\bigg] \\
    & + \sigma^2\delta_{ij}.
\end{aligned}
\end{equation*}

Astrophysical interpretations of the \celerite kernel hyperparameters are analogous to those of the quasi-periodic kernel.  $B$ is the amplitude, $L$ is the decay timescale for the exponential term, $P_{\textrm{rot}}$ is the recurrence timescale for the periodic term (i.e.~the stellar rotation period), and $C$ is a scaling term.  $\sigma^2\delta_{ij}$ is a ``jitter" term that accounts for additional white noise not accounted for by the formal errors on the data.  While these interpretations are useful for providing a more intuitive understanding of the GP model, \citet{angus2018} caution that the hyperparameters other than $P_{\textrm{rot}}$ are often degenerate, and the short time baseline of the TESS lightcurve prevent us from obtaining tight constraints on them.  In particular, since we see no evolution of the phase curve during TESS observations, the parameters dealing with signal decay are difficult to fit.  Thus, while we are confident in our determination of the rotation period, a more quantitative analysis of spot lifetime and decay for G 227-22 would require longer-baseline space photometry.

We modeled the TESS lightcurve of G 227-22 as a \celerite GP using the \emcee MCMC ensemble sampler \citep{dfm2013}.  We binned the sigma-clipped lightcurve by a factor of 10, so that each binned point represents a time step of 20 minutes.  Table \ref{tab:g227-22_mcmc} shows the priors adopted for our model; we constrained $P_{\textrm{rot}}$ based on the period from \citet{newton2016} and our preliminary analysis of the TESS lightcurve, and placed uninformative log-uniform priors on the other hyperparameters.  Our MCMC run used 50 random walkers, initialized with small perturbations from a maximum-likelihood initial fit.  The MCMC chains were allowed to proceed for up to $10^5$ steps, although we halted the calculation when the chains converged, where convergence is defined as having a Gelman-Rubin statistic value within 0.1\% of unity \citep{ford2006}.  The results of our model are listed in Table \ref{tab:g227-22_mcmc}. Our adopted values and their uncertainties are taken from the median of the posterior probability distribution and the 16th/84th percentiles of the distribution, respectively.  In the case of the rotation period, we obtain identical values by fitting a normal distribution to the posterior, as shown in Figure \ref{fig:g227-22_rotation}c.

\begin{table}
\begin{center}
\begin{tabular}{| l c  c |}

\multicolumn{3}{ c }{GP Model Hyperparameters for G 227-22 Lightcurve} \\
\hline
Parameter & Prior & Posterior \\
\hline
$\log B$ & $\mathscr{U}(-6, 0)$ & $-1.4^{+0.5}_{-0.4}$\\
$\log L$ (d) & $\mathscr{U}(-6, 6)$ & $3.8^{+0.5}_{-0.4}$ \\
$\log C$ & $\mathscr{U}(-6, 6)$ & $4.2 \pm 0.2$ \\
$P_{\textrm{rot}}$ (d) & $\mathscr{N}(0.28, 0.03)$ & $0.28018 \pm 1.0 \times 10^{-5}$\\
$\sigma$ & $\mathscr{U}(10^{-9}, 10^{-3})$ & $4.5 \times 10^{-4} \pm 1.3 \times 10^{-5}$ \\
\hline
\end{tabular}
\caption{Priors and 1-dimensional posterior distributions for the hyperparameters of our GP model to the G 227-22 TESS lightcurve.}
\label{tab:g227-22_mcmc}
\end{center}
\end{table}

In Figure \ref{fig:g227-22_rotation}, we show the results of our GP model for a representative chunk of the data.  We also show the stability of the signal's amplitude and phase from two perspectives.  First, we fold all data to the best-fit period.  Then, we divide the lightcurve into 7-day chunks and bin it in phase space, taking the median flux in each bin to minimize the impact of flares.  We show the stacked phase curves \citep[as shown in, e.g.,][]{lurie2015,davenport2015} to elucidate potential drifts in the phase curve.  We find that the phase and amplitude of the rotational modulation is extremely stable, resulting in tight constraints on the period; our $1\sigma$ uncertainty is just under 1 second.

\subsubsection{RV modulation}
Both HPF and HIRES RVs of G 227-22 show clear, approximately sinusoidal variability at the 6-hour rotation period.  The amplitude of the modulation in the HPF time series is approximately 140 \unit{m/s}, whereas the HIRES RVs show a much higher amplitude of 244 \unit{m/s}.  During our HIRES observations, the star underwent a minor flare, as evidenced by emission in several of the spectrum's He I lines.  The RVs from these spectra are anomalously blueshifted from the expected peak of the rotation phase curve, potentially due to the geometry of the flare relative to the rotationally Doppler-shifted stellar photosphere \citep[e.g.][]{reiners2009}.  As observed for GJ 1245B, the HPF RVs of G 227-22 show rotational modulation that appears to remain coherent over a single observing season.  Our HIRES observations took place over a single night, so those RVs alone do not inform us as to the signal's longevity.  

\begin{figure}[h]
    \centering
    \includegraphics[width=\columnwidth]{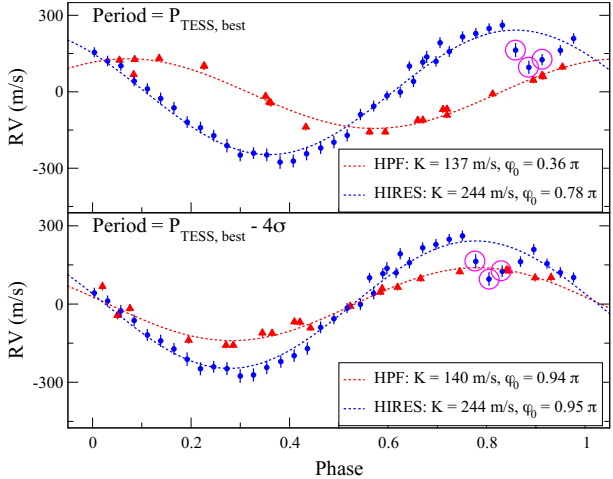}
    \caption{HPF and HIRES RVs of G 227-22, folded to the stellar rotation period.  HIRES RVs taken during the stellar flare are marked with pink circles.  In the top panel, we have used the best-fit period derived from the TESS lightcurve, while the lower panel uses a period that is $4\sigma$ shorter.  The phase ($\phi_0$) of the RV signals is consistent when we adopt the shorter period.}
    \label{fig:g227-22_rv}
\end{figure}

Given the stability of the photometric rotation signal across the TESS lightcurve, we sought to determine whether the RVs were consistent with a coherent signal across the entire $\sim1$-year observational baseline.  In the top panel of Figure \ref{fig:g227-22_rv}, we show the RVs folded to the best-fit 0.28018-day rotation period derived from the TESS lightcurve.  At that period, there is a large phase shift between the 2018 HPF and 2019 HIRES RV signals.  However, if the true period is shorter than our adopted value by 4 standard deviations, the RVs are approximately in phase.  The lower panel of Figure \ref{fig:g227-22_rv} shows the RVs folded to the shorter period.

We list three possible interpretations of these results:

\begin{enumerate}

    \item The 0.28018-day rotation period is correct, and the RV signal has changed phase between the two observing seasons.  The phase shift may be due to spots decaying and reappearing in different locations, or because of differential rotation shifting spots' relative longitudes.
    
    \item The shorter 0.28014-day rotation period is correct, and the RV signals are in phase.  In this case, it is likely that the \textbf{spot configurations} producing the RV signals and the photometric phase curve are the same, and have remained mostly unchanged over at least 23 months, or approximately 2500 stellar rotations.
    
    \item The TESS lightcurve samples a different stellar latitude than the RVs.  These latitudes rotate differentially, but their atmospheric features are long-lived.  In this case, the photometric and RV signals may both be in phase with themselves, but do not have exactly the same period.  We consider this to be the least likely possibility, since the stability of the TESS rotation signal would appear to be inconsistent with the presence of differential rotation.
    
\end{enumerate}

\subsection{GJ 3959}

\subsubsection{Determining the rotation period}

\begin{figure}
    \centering
    \includegraphics[width=\columnwidth]{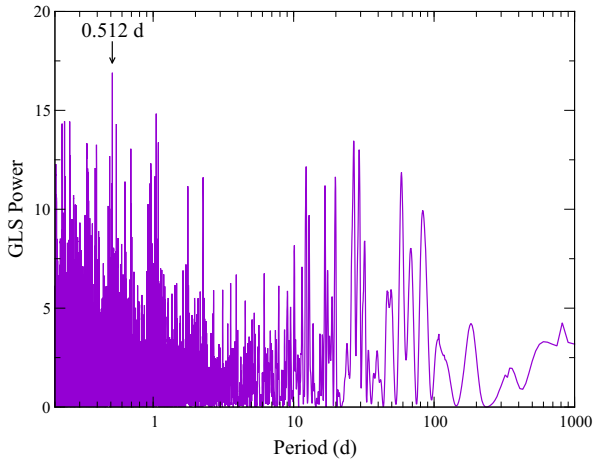}
    \caption{Generalized Lomb-Scargle periodogram of our LCO $V$-band photometry of GJ 3959.  The highest peak occurs at 0.512 days, matching the rotation period determined by \citet{newton2016}.}
    \label{fig:gj3959_ps}
\end{figure}

As of this writing, TESS lightcurves for GJ 3959 are not yet available, so we must rely on ground-based photometric resources to determine its rotation period.  \citet{newton2016} found evidence for a period of 0.512 days based on photometry from MEarth.  We have observed GJ 3959 using the LCO network over 2.5 years from 2017-2020, amassing 487 usable images in the \iprime filter.  In the $V$ band, we used 3 exposures per visit because of the star's relative faintness in this band, and therefore acquired 1266 usable images.  Given the relatively short candidate rotation period, we have not binned our $V$-band data.  In Figure \ref{fig:gj3959_ps}, we show a generalized Lomb-Scargle \citep[GLS;][]{zk2009} periodogram of our $V$-band time series, which shows a marginally-significant peak at the 0.512-day period proposed by \citet{newton2016}.  The corresponding \iprime periodogram shows no significant peaks other than those created by our daily sampling cadence, probably because of the reduced amplitude of the photometric rotation signal at longer wavelengths.

We find that neither the MEarth nor the LCO photometry have both the precision and cadence necessary to confidently constrain the long-term coherence of the photometric phase curve.  However, given the agreement between the power spectra of the LCO and MEarth lightcurves on the 0.512d period, and the subsequent agreement of our RVs with this period (described in the next subsection), we have adopted 0.51207 days as the rotation period of GJ 3959.

\subsection{RV analysis}

The rotation curve of GJ 3959 is especially difficult to sample from ground-based facilities because of its proximity to half of an Earth day.  The period is too long to sample at low airmass over a single night.  For queue-scheduled observations with HET, the fixed-altitude design of the telescope limits observability to certain ``tracks" within a night \citep{shetrone2007}, so nightly observations typically result in repeated visits in the same region of phase space.  As a result, our RV sampling of the rotational modulation for GJ 3959 is poorer than for the other targets presented here.  For HPF, we have obtained 12 RVs between January and August of 2019, while our HIRES RVs comprise 15 velocities taken across May 17 and June 8 of 2019.

\begin{figure}
    \centering
    \includegraphics[width=\columnwidth]{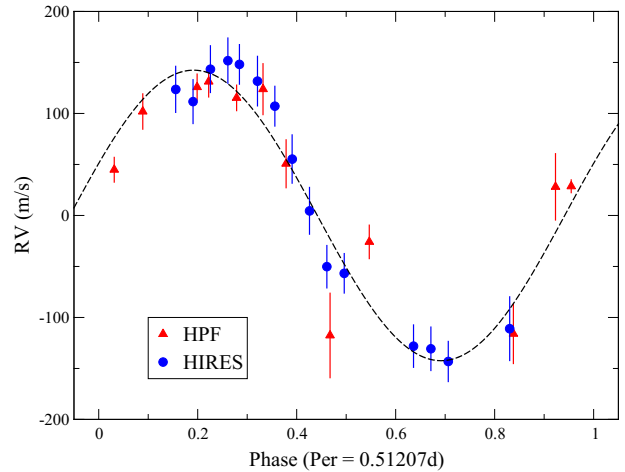}
    \caption{HPF and HIRES RVs of GJ 3959, folded to the stellar rotation period.  A zero-point offset between the spectrometers has been applied.  A sinusoid fit to the combined RV series is shown as a dashed line.}
    \label{fig:gj3959_rv}
\end{figure}

Our HIRES RVs again show a coherent, approximately sinusoidal signal at the stellar rotation period, and are amenable to fits at several periods near 0.5 days.  The HPF RVs, on the other hand, can only be well modeled with a period very close to the 0.512-day period preferred by the LCO and MEarth lightcurves.  In Figure \ref{fig:gj3959_rv}, we show all RVs phase-folded to the 0.512d period.  The RVs appear remarkably consistent with a coherent, achromatic signal, although neither the HPF nor the HIRES time series have sufficient phase coverage to model the signal separately to determine their consistency.  Still, given the typical factor of $\sim2$ difference in RV amplitude observed between the optical and NIR signals for AD Leo, G 227-22 and (sometimes) GJ 1245B, it is somewhat unexpected to see such similar levels of modulation for both data sets.  Given the phase coherence of the signals, and the relative proximity in time of the observations, it seems unlikely that the optical and NIR RVs sample completely different starspot distributions, as seen for GJ 1245B.

Given the low amplitude of the photometric rotation signal, we find that we can best constrain the stellar rotation period using our RV data.  The $0.51207 \pm 5 \times 10^{-5}$-day period listed in Table \ref{tab:stellar} is based on a sinusoidal fit of the combined HIRES/HPF RV series using \radvelp.  If we instead model only the HIRES RVs, we obtain a period of $0.51336 \pm 8 \times 10^{-4}$ day, which is consistent with our adopted value to less than $2 \sigma$.  The HPF RVs alone do not provide sufficient phase coverage to constrain a model.  As Figure \ref{fig:gj3959_rv} shows, only the combined data covers the full phase space of the rotation signal, which motivates our choice to adopt the period from the model to both data sets.

\citet{talor2018} showed 12 RVs of GJ 3959 from the visible-channel CARMENES spectrometer.  The CARMENES RVs show strong anticorrelation with the chromatic index \citep[CRX;][]{zechmeister2018}, suggesting the Doppler variability \emph{is} wavelength dependent.  Our HPF RVs show no such dependence on CRX, suggesting a flattening of the dependence at NIR wavelengths.  We do note that our current RV pipeline for HPF only extracts velocities from orders covering wavelengths from $850-1080$ nm, whereas the visible CARMENES RVs are extracted across roughly $400$ nm.  Thus, the CARMENES CRX metric offers more leverage for evaluating chromatic dependence.  Because the HIRES RVs are extracted from a relatively narrow range of wavelengths bounded by the I$_2$ absorption band, no such chromatic dependence can be evaluated.  Given the phase coherence of the HIRES and HPF RVs, we conclude that the most likely explanation for the matching optical/NIR amplitudes is that a single spot or spot complex decreased its filling factor between the HPF and HIRES observations such that the amplitudes approximately match.  Another possibility is that the chromatic dependence of the rotation signal's amplitude is not monotonic, and increases for certain NIR wavelengths.  As suggested by \citet{reiners2013}, this may be due to Zeeman splitting in the NIR.  In either case, the chromaticity claimed by \citet{talor2018} suggests that the observed RV periodicity is again caused by starspot activity rather than a spin-orbit coupled exoplanet, despite the agreement between optical and NIR RVs observed here.

\section{Discussion}
\label{sec:discussion}

We have found that large-amplitude RV signals at the rotation periods of rapidly-rotating M dwarfs is common.  This result, at its most basic level, is unsurprising; rapidly rotating dwarfs of all spectral types are known to exhibit activity-induced RV signals \citep[e.g.,][]{queloz2001,haywood2014,zechmeister2018}.  What is most interesting about our results are the amplitudes and persistence of the RV rotation signals for these M stars.  In the case of G 227-22, a starspot signal with just 0.2\% amplitude---which is extremely difficult to measure from ground-based facilities---gives rise to an RV signal hundreds of meters per second in amplitude.  Low photometric variability is known to sometimes belie large spot filling factors in young stars \citep[e.g.][]{gully2017}, which probably explains the discrepancy between the Doppler and photometric amplitudes observed herein.  We find that these RV amplitudes are frequently lower in the NIR, but not reliably so.  While our sample is admittedly small, we find one example---GJ 3959---where the NIR amplitude is comparable to the optical, at least over a period of 6 months to a year.  While GJ 1245B also shows a NIR amplitude comparable to a single season (2010) from HIRES, \kep data from 2010 suggests the star was especially quiet for that season, complicating the direct comparison between 2010 and 2018 RVs.

For all of our targets, we find that it is typical for rotationally-modulated RV signals to remain coherent for at least one full observing season.  At these rotation periods, such coherence implies that the overall distribution of magnetically-induced surface inhomogeneities---if not individual starspots or spot complexes themselves---remains consistent for hundreds of stellar rotations, at minimum.  This is a dramatic departure from Sunlike stars, for which spots tend to decay over 2-3 stellar rotations \citep[e.g.,][]{lopezmorales2016}.  For the Sun specifically, spot lifetimes are strongly correlated with size \citep[e.g.][]{martinezpillet1993}, and most spots disappear in less than a single rotation.

The examples of GJ 1245B \citep{lurie2015}, GJ 1243 \citep{davenport2015}, and G 227-22 (\S~\ref{sec:g227-22}) provide useful context in which to place our RV observations.  For each of these rapidly-rotating M dwarfs, space-based photometers show phase curves which remain coherent over typical timescales of 6 months to 1 year before evolving.  This evolution timescale is well-matched to an observing season from a ground-based facility, and would explain why the RV signals remain coherent for as long as they do.  \citet{lurie2015} suggest that differential rotation may be responsible for the phase shifts observed for GJ 1245B, as spot complexes change their relative longitudinal separation over time.  Several starspot signals observed in RV have shown evidence of differential rotation \citep{santos2014,johnson2016}, again suggesting consistency with photometric observations.  Assuming this physical model is applicable, it was probably not necessary for us to select targets with low stellar inclinations; long-lived rotation signals should be present for rapidly-rotating stars at a range of inclination values.

We find that the joint HIRES+HARPS+HPF RV time series for AD Leo is incompatible with the presence of a giant exoplanet in spin-orbit resonance, in agreement with \citet{carleo2020}.  The starspot signal that caused \citet{tuomi2018} to propose a planet persisted throughout the 6-month observing season in 2006.  This behavior is inconsistent with the rapid phase evolution of the MOST lightcurve from \citet{huntwalker2012}, which shows the photometric signal evolving over approximately 2 cycles.  \citet{carleo2020} suggest that instead of originating from dark starspots, the RV variations for AD Leo might be tied to inhibition of convection from the global magnetic field.  In this hypothesis, the RV rotation signal will decay as the global field evolves \citep[as observed by][]{lavail2018}, rather than with individual starspots.  Further evidence that the RV rotation signal may not be entirely due to starspots is the 8 \unit{m/s} NIR amplitude we observe with HPF.  This amplitude is consistent with the prediction of \citet{reiners2013} and \citet{carleo2020} if the NIR variability is caused by Zeeman splitting rather than spot-photosphere contrast.

On the other hand, the RVs of GJ 1245B and G 227-22 appear to be much more consistent with starspot modulation.  Changes in amplitude and phase of the GJ 1245B match changes in the \kep lightcurve, and the consistency of the G 227-22 RV curve is matched by the TESS lightcurve.  We posit that these more rapidly-rotating stars should have more powerful magnetic fields, which may suppress differential rotation and starspot evolution \citep[as discussed in][]{kuker2005,lurie2015,davenport2015,shulyak2017}.  If so, we should expect both RV and photometric time series to evolve more slowly, as observed for GJ 1245B and G 227-22.

More broadly, our results suggest that confidently detecting exoplanets at orbital periods near the stellar rotation period or its harmonics may be even more difficult than previously appreciated.  The community has often sought to discriminate true exoplanets from false positives by relying on the expectations that 1) an exoplanet signal will remain coherent indefinitely, while activity signals decay, and 2) exoplanet signals are achromatic, while activity signals decrease in amplitude with increasing wavelength.  In this study, we show examples of starspot signals that violate both of these assumptions.  Furthermore, there is reason to suspect that starspot signals for M dwarfs may survive for many rotations even as the stars age and spin down.  For example, the rotation signal for Kapteyn's star \citep{robertson2015} appears to remain in phase over 10 years of HARPS and HIRES observations.  Thus, any potential exoplanet signal near an activity-induced periodicity will need to be carefully vetted using additional information besides coherence and chromaticity.

On the other hand, the regularity of the observed rotation signals---at least across a single observing season---potentially creates an opportunity for detecting exoplanets orbiting young M dwarfs.  If the rotation signal does not evolve over an observing season, it should be possible to model and remove it, revealing exoplanet signals in the residuals.  This is especially true if contemporaneous space-based photometry is available to reliably model the activity signal.  The precision and cadence of TESS photometry for G 227-22 allowed us to constrain its rotation period extremely precisely; our 0.9-second uncertainty on the period is, to our knowledge, the smallest uncertainty for the rotation period of a main-sequence star other than the Sun.  For comparison, the most precisely determined periods from the large autocorrelation analysis of \citet{mcquillan2014} have typical uncertainties of a few hundred seconds, although here we are admittedly comparing the bespoke analysis of a single star to batch analysis of thousands.  A more apt comparison might be to GJ 1243, the period of which \citet{davenport2015} determined to a $1\sigma$ uncertainty of about 18 seconds.  Furthermore, at these precisions, we should be explicit that we are likely measuring the period of a single spot or spot complex, and that the bulk rotation of the star may be more complex than can be expressed as a fixed period and uncertainty.  For other M dwarfs exhibiting similarly coherent spot signals, it should be possible to determine equally precise variability periods and model RVs to high accuracy.

As detailed in \S\ref{sec:targets}, our results are based on a small sample of young, low-mass M stars.  Thus, the exact degree to which the phenomena observed herein can be generalized to all Doppler survey targets is unclear.  However, the large and diverse data sets available for these targets allowed us to investigate in greater detail what has been recently observed in other studies of larger numbers of objects \citep[e.g.][]{suarez2018,talor2018}.  Namely, that starspot signals on M dwarfs can exhibit markedly different characteristics from those on hotter stars, and that the propagation of those signals to RV is not reliably predictable.

\section{Summary and Conclusion}
\label{sec:summary}
We have obtained highly precise optical and NIR RVs of four rapidly-rotating M dwarfs using Keck/HIRES and HET/HPF.  All four stars exhibit large-amplitude RV signals at their rotation periods.  These signals remain consistent in amplitude and phase over typical timescales of at least a single observing season.  In general, the NIR amplitudes of these signals are lower than the optical, but not predictably so, and we identify one star---GJ 3959---for which we cannot clearly establish a difference between the optical and NIR RV signals.  The persistence and unpredictable chromaticity of these signals emphasizes the challenge of detecting true exoplanet signals near the rotation period or its harmonics/aliases with RV.

\acknowledgments
The authors thank the anonymous referee for a prompt and thorough review.  Their comments significantly improved the quality of this manuscript.

This work was partially supported by funding from the Center for Exoplanets and Habitable Worlds. The Center for Exoplanets and Habitable Worlds is supported by the Pennsylvania State University, the Eberly College of Science, and the Pennsylvania Space Grant Consortium. This work was supported by NASA Headquarters under the NASA Earth and Space Science Fellowship Program through grants NNX16AO28H and 80NSSC18K1114. We acknowledge support from NSF grants AST-1006676, AST-1126413, AST-1310875, AST-1310885, AST-1517592, the NASA Astrobiology Institute (NAI; NNA09DA76A), and PSARC in our pursuit of precision radial velocities in the NIR. Computations for this research were performed on the Pennsylvania State University’s Institute for Computational and Data Sciences. We acknowledge support from the Heising-Simons Foundation via grant 2017-0494 and 2019-1177. This is University of Texas Center for Planetary Systems Habitability Contribution 0003.

We wish to thank Kento Masuda for informative discussions on determining stellar inclinations from projected rotational velocities, stellar radii, and rotation periods.

These results are based on observations obtained with the Habitable-zone Planet Finder Spectrograph on the Hobby-Eberly Telescope. We thank the Resident astronomers and Telescope Operators at the HET for the skillful execution of our observations of our observations with HPF. The Hobby-Eberly Telescope is a joint project of the University of Texas at Austin, the Pennsylvania State University, Ludwig-Maximilians-Universität München, and Georg-August Universität Gottingen. The HET is named in honor of its principal benefactors, William P. Hobby and Robert E. Eberly. The HET collaboration acknowledges the support and resources from the Texas Advanced Computing Center. 

Some of the data presented herein were obtained at the W. M. Keck Observatory, which is operated as a scientific partnership among the California Institute of Technology, the University of California and the National Aeronautics and Space Administration. The Observatory was made possible by the generous financial support of the W. M. Keck Foundation. The authors wish to recognize and acknowledge the very significant cultural role and reverence that the summit of Maunakea has always had within the indigenous Hawaiian community.  We are most fortunate to have the opportunity to conduct observations from this mountain. The authors also thank the Keck Observing Assistants and Telescope Operators for their assistance in collecting these data.

This research has made use of the Keck Observatory Archive (KOA), which is operated by the W. M. Keck Observatory and the NASA Exoplanet Science Institute (NExScI), under contract with the National Aeronautics and Space Administration.

This work makes use of observations from the LCOGT network through the observing program KEY2017AB-005.

This paper includes data collected by the TESS mission. Funding for the TESS mission is provided by the NASA Explorer Program. This work has made use of data from the European Space Agency (ESA) mission {\it Gaia} (\url{https://www.cosmos.esa.int/gaia}), processed by the {\it Gaia} Data Processing and Analysis Consortium (DPAC, \url{https://www.cosmos.esa.int/web/gaia/dpac/consortium}). Funding for the DPAC has been provided by national institutions, in particular the institutions participating in the {\it Gaia} Multilateral Agreement.

\facilities{HET, Keck:I (HIRES), LCOGT, TESS} 

Software:
\texttt{AstroImageJ} \citep{collins2017}, 
\texttt{Astropy} \citep{astropy2013},
\texttt{Astroquery} \citep{ginsburg2016},
\texttt{celerite} \citep{dfm2017},
\texttt{corner.py} \citep{dfm2016}, 
\texttt{emcee} \citep{dfm2013},
\texttt{HxRGproc} \citep{ninan2018},
\texttt{Isochrones} \citep{morton2015},
\texttt{Jupyter} \citep{jupyter2016},
\texttt{matplotlib} \citep{hunter2007},
\texttt{numpy} \citep{vanderwalt2011},
\texttt{pandas} \citep{pandas2010},
\texttt{radvel} \citep{fulton2018},
\texttt{SERVAL} \citep{zechmeister2018}

\pagebreak
\bibliographystyle{yahapj}
\bibliography{references}

\begin{thebibliography}{}
\providecommand\natexlab[1]{#1}
\providecommand\JournalTitle[1]{#1}

\bibitem[{{Ambikasaran} {et~al.}(2015){Ambikasaran}, {Foreman-Mackey},
  {Greengard}, {Hogg}, \& {O'Neil}}]{ambikasaran15}
{Ambikasaran}, S., {Foreman-Mackey}, D., {Greengard}, L., {Hogg}, D.~W., \&
  {O'Neil}, M. 2015,
  \href{http://dx.doi.org/10.1109/TPAMI.2015.2448083}{\JournalTitle{IEEE
  Transactions on Pattern Analysis and Machine Intelligence}, 38, 252}

\bibitem[{{Angus} {et~al.}(2018){Angus}, {Morton}, {Aigrain}, {Foreman-Mackey},
  \& {Rajpaul}}]{angus2018}
{Angus}, R., {Morton}, T., {Aigrain}, S., {Foreman-Mackey}, D., \& {Rajpaul},
  V. 2018,
  \href{http://dx.doi.org/10.1093/mnras/stx2109}{\JournalTitle{\mnras}, 474,
  2094}

\bibitem[{{Astropy Collaboration} {et~al.}(2013){Astropy Collaboration},
  {Robitaille}, {Tollerud}, {Greenfield}, {Droettboom}, {Bray}, {Aldcroft},
  {Davis}, {Ginsburg}, {Price-Whelan}, {Kerzendorf}, {Conley}, {Crighton},
  {Barbary}, {Muna}, {Ferguson}, {Grollier}, {Parikh}, {Nair}, {Unther},
  {Deil}, {Woillez}, {Conseil}, {Kramer}, {Turner}, {Singer}, {Fox}, {Weaver},
  {Zabalza}, {Edwards}, {Azalee Bostroem}, {Burke}, {Casey}, {Crawford},
  {Dencheva}, {Ely}, {Jenness}, {Labrie}, {Lim}, {Pierfederici}, {Pontzen},
  {Ptak}, {Refsdal}, {Servillat}, \& {Streicher}}]{astropy2013}
{Astropy Collaboration}, {Robitaille}, T.~P., {Tollerud}, E.~J., {et~al.} 2013,
  \href{http://dx.doi.org/10.1051/0004-6361/201322068}{\JournalTitle{AAP}, 558,
  A33}

\bibitem[{{Bailer-Jones} {et~al.}(2018){Bailer-Jones}, {Rybizki}, {Fouesneau},
  {Mantelet}, \& {Andrae}}]{bailerjones2018}
{Bailer-Jones}, C.~A.~L., {Rybizki}, J., {Fouesneau}, M., {Mantelet}, G., \&
  {Andrae}, R. 2018,
  \href{http://dx.doi.org/10.3847/1538-3881/aacb21}{\JournalTitle{\aj}, 156,
  58}

\bibitem[{{Boisse} {et~al.}(2011){Boisse}, {Bouchy}, {H{\'e}brard}, {Bonfils},
  {Santos}, \& {Vauclair}}]{boisse2011}
{Boisse}, I., {Bouchy}, F., {H{\'e}brard}, G., {et~al.} 2011,
  \href{http://dx.doi.org/10.1051/0004-6361/201014354}{\JournalTitle{\aap},
  528, A4}

\bibitem[{{Borucki} {et~al.}(2010){Borucki}, {Koch}, {Basri}, {Batalha},
  {Brown}, {Caldwell}, {Caldwell}, {Christensen-Dalsgaard}, {Cochran},
  {DeVore}, {Dunham}, {Dupree}, {Gautier}, {Geary}, {Gilliland}, {Gould},
  {Howell}, {Jenkins}, {Kondo}, {Latham}, {Marcy}, {Meibom}, {Kjeldsen},
  {Lissauer}, {Monet}, {Morrison}, {Sasselov}, {Tarter}, {Boss}, {Brownlee},
  {Owen}, {Buzasi}, {Charbonneau}, {Doyle}, {Fortney}, {Ford}, {Holman},
  {Seager}, {Steffen}, {Welsh}, {Rowe}, {Anderson}, {Buchhave}, {Ciardi},
  {Walkowicz}, {Sherry}, {Horch}, {Isaacson}, {Everett}, {Fischer}, {Torres},
  {Johnson}, {Endl}, {MacQueen}, {Bryson}, {Dotson}, {Haas}, {Kolodziejczak},
  {Van Cleve}, {Chandrasekaran}, {Twicken}, {Quintana}, {Clarke}, {Allen},
  {Li}, {Wu}, {Tenenbaum}, {Verner}, {Bruhweiler}, {Barnes}, \&
  {Prsa}}]{borucki2010}
{Borucki}, W.~J., {Koch}, D., {Basri}, G., {et~al.} 2010,
  \href{http://dx.doi.org/10.1126/science.1185402}{\JournalTitle{Science}, 327,
  977}

\bibitem[{{Brown} {et~al.}(2011){Brown}, {Miesch}, {Browning}, {Brun}, \&
  {Toomre}}]{brown2011}
{Brown}, B.~P., {Miesch}, M.~S., {Browning}, M.~K., {Brun}, A.~S., \& {Toomre},
  J. 2011,
  \href{http://dx.doi.org/10.1088/0004-637X/731/1/69}{\JournalTitle{\apj}, 731,
  69}

\bibitem[{{Brown} {et~al.}(2013){Brown}, {Baliber}, {Bianco}, {Bowman},
  {Burleson}, {Conway}, {Crellin}, {Depagne}, {De Vera}, {Dilday}, {Dragomir},
  {Dubberley}, {Eastman}, {Elphick}, {Falarski}, {Foale}, {Ford}, {Fulton},
  {Garza}, {Gomez}, {Graham}, {Greene}, {Haldeman}, {Hawkins}, {Haworth},
  {Haynes}, {Hidas}, {Hjelstrom}, {Howell}, {Hygelund}, {Lister}, {Lobdill},
  {Martinez}, {Mullins}, {Norbury}, {Parrent}, {Paulson}, {Petry}, {Pickles},
  {Posner}, {Rosing}, {Ross}, {Sand}, {Saunders}, {Shobbrook}, {Shporer},
  {Street}, {Thomas}, {Tsapras}, {Tufts}, {Valenti}, {Vander Horst}, {Walker},
  {White}, \& {Willis}}]{brown2013}
{Brown}, T.~M., {Baliber}, N., {Bianco}, F.~B., {et~al.} 2013,
  \href{http://dx.doi.org/10.1086/673168}{\JournalTitle{\pasp}, 125, 1031}

\bibitem[{{Buccino} {et~al.}(2014){Buccino}, {Petrucci}, {Jofr{\'e}}, \&
  {Mauas}}]{buccino2014}
{Buccino}, A.~P., {Petrucci}, R., {Jofr{\'e}}, E., \& {Mauas}, P.~J.~D. 2014,
  \href{http://dx.doi.org/10.1088/2041-8205/781/1/L9}{\JournalTitle{\apjl},
  781, L9}

\bibitem[{{Butler} {et~al.}(1996){Butler}, {Marcy}, {Williams}, {McCarthy},
  {Dosanjh}, \& {Vogt}}]{butler1996}
{Butler}, R.~P., {Marcy}, G.~W., {Williams}, E., {et~al.} 1996,
  \href{http://dx.doi.org/10.1086/133755}{\JournalTitle{\pasp}, 108, 500}

\bibitem[{{Butler} {et~al.}(2017){Butler}, {Vogt}, {Laughlin}, {Burt},
  {Rivera}, {Tuomi}, {Teske}, {Arriagada}, {Diaz}, {Holden}, \&
  {Keiser}}]{butler2017}
{Butler}, R.~P., {Vogt}, S.~S., {Laughlin}, G., {et~al.} 2017,
  \href{http://dx.doi.org/10.3847/1538-3881/aa66ca}{\JournalTitle{\aj}, 153,
  208}

\bibitem[{{Carleo} {et~al.}(2020){Carleo}, {Malavolta}, {Lanza}, {Damasso},
  {Desidera}, {Borsa}, {Mallonn}, {Pinamonti}, {Gratton}, {Alei}, {Benatti},
  {Mancini}, {Maldonado}, {Biazzo}, {Esposito}, {Frustagli},
  {Gonz{\'a}lez-{\'A}lvarez}, {Micela}, {Scandariato}, {Sozzetti}, {Affer},
  {Bignamini}, {Bonomo}, {Claudi}, {Cosentino}, {Covino}, {Fiorenzano},
  {Giacobbe}, {Harutyunyan}, {Leto}, {Maggio}, {Molinari}, {Nascimbeni},
  {Pagano}, {Pedani}, {Piotto}, {Poretti}, {Rainer}, {Redfield}, {Baffa},
  {Baruffolo}, {Buchschacher}, {Billotti}, {Cecconi}, {Falcini}, {Fantinel},
  {Fini}, {Galli}, {Ghedina}, {Ghinassi}, {Giani}, {Gonzalez}, {Gonzalez},
  {Guerra}, {Hernandez Diaz}, {Hernandez}, {Iuzzolino}, {Lodi}, {Oliva},
  {Origlia}, {Perez Ventura}, {Puglisi}, {Riverol}, {Riverol}, {San Juan},
  {Sanna}, {Scuderi}, {Seemann}, {Sozzi}, \& {Tozzi}}]{carleo2020}
{Carleo}, I., {Malavolta}, L., {Lanza}, A.~F., {et~al.} 2020,
  \JournalTitle{arXiv e-prints}, arXiv:2002.10562

\bibitem[{{Chabrier} \& {Baraffe}(1997)}]{chabrier1997}
{Chabrier}, G., \& {Baraffe}, I. 1997, \JournalTitle{\aap}, 327, 1039

\bibitem[{{Ciardi} {et~al.}(2015){Ciardi}, {van Eyken}, {Barnes}, {Beichman},
  {Carey}, {Crockett}, {Eastman}, {Johns-Krull}, {Howell}, {Kane}, {. Mclane},
  {Plavchan}, {Prato}, {Stauffer}, {van Belle}, \& {von Braun}}]{ciardi2015b}
{Ciardi}, D.~R., {van Eyken}, J.~C., {Barnes}, J.~W., {et~al.} 2015,
  \href{http://dx.doi.org/10.1088/0004-637X/809/1/42}{\JournalTitle{\apj}, 809,
  42}

\bibitem[{{Collins} {et~al.}(2017){Collins}, {Kielkopf}, {Stassun}, \&
  {Hessman}}]{collins2017}
{Collins}, K.~A., {Kielkopf}, J.~F., {Stassun}, K.~G., \& {Hessman}, F.~V.
  2017, \JournalTitle{ArXiv e-prints},
  \href{http://arxiv.org/abs/1701.04817}{{\sffamily arXiv:1701.04817
  [astro-ph.IM]}}

\bibitem[{{Crockett} {et~al.}(2012){Crockett}, {Mahmud}, {Prato},
  {Johns-Krull}, {Jaffe}, {Hartigan}, \& {Beichman}}]{crockett2012}
{Crockett}, C.~J., {Mahmud}, N.~I., {Prato}, L., {et~al.} 2012,
  \href{http://dx.doi.org/10.1088/0004-637X/761/2/164}{\JournalTitle{\apj},
  761, 164}

\bibitem[{{Cutri} {et~al.}(2003){Cutri}, {Skrutskie}, {van Dyk}, {Beichman},
  {Carpenter}, {Chester}, {Cambresy}, {Evans}, {Fowler}, {Gizis}, {Howard},
  {Huchra}, {Jarrett}, {Kopan}, {Kirkpatrick}, {Light}, {Marsh}, {McCallon},
  {Schneider}, {Stiening}, {Sykes}, {Weinberg}, {Wheaton}, {Wheelock}, \&
  {Zacarias}}]{cutri2003}
{Cutri}, R.~M., {Skrutskie}, M.~F., {van Dyk}, S., {et~al.} 2003,
  \JournalTitle{VizieR Online Data Catalog}, II/246

\bibitem[{{Dalba} {et~al.}(2020){Dalba}, {Gupta}, {Rodriguez}, {Dragomir},
  {Huang}, {Kane}, {Quinn}, {Bieryla}, {Esquerdo}, {Fulton}, {Scarsdale},
  {Batalha}, {Crossfield}, {Dressing}, {Howard}, {Huber}, {Isaacson},
  {Petigura}, {Robertson}, {Roy}, {Weiss}, {Knudstrup}, {Andersen}, {Grundahl},
  {Yao}, {Pepper}, {Villanueva Jr.}, {Ciardi}, {Cloutier}, {Jacobs},
  {Kristiansen}, {LaCourse}, {Lendl}, {Osborn}, {Palle}, {Stassun}, {Stevens},
  {Ricker}, {Vanderspek}, {Latham}, {Seager}, {Winn}, {Jenkins}, {Caldwell},
  {Daylan}, {Fong}, {Goeke}, {Rose}, {Rowden}, {Schlieder}, {Smith}, \&
  {Vanderburg}}]{dalba2020}
{Dalba}, P.~A., {Gupta}, A.~F., {Rodriguez}, J.~E., {et~al.} 2020,
  \JournalTitle{arXiv e-prints}, arXiv:2003.10451

\bibitem[{{Davenport} {et~al.}(2015){Davenport}, {Hebb}, \&
  {Hawley}}]{davenport2015}
{Davenport}, J. R.~A., {Hebb}, L., \& {Hawley}, S.~L. 2015,
  \href{http://dx.doi.org/10.1088/0004-637X/806/2/212}{\JournalTitle{\apj},
  806, 212}

\bibitem[{{Delfosse} {et~al.}(1998){Delfosse}, {Forveille}, {Perrier}, \&
  {Mayor}}]{delfosse1998}
{Delfosse}, X., {Forveille}, T., {Perrier}, C., \& {Mayor}, M. 1998,
  \JournalTitle{\aap}, 331, 581

\bibitem[{{Delfosse} {et~al.}(2000){Delfosse}, {Forveille}, {S{\'e}gransan},
  {Beuzit}, {Udry}, {Perrier}, \& {Mayor}}]{delfosse2000}
{Delfosse}, X., {Forveille}, T., {S{\'e}gransan}, D., {et~al.} 2000,
  \JournalTitle{\aap}, 364, 217

\bibitem[{{Dressing} \& {Charbonneau}(2015)}]{dressing2015}
{Dressing}, C.~D., \& {Charbonneau}, D. 2015,
  \href{http://dx.doi.org/10.1088/0004-637X/807/1/45}{\JournalTitle{\apj}, 807,
  45}

\bibitem[{{Endl} {et~al.}(2006){Endl}, {Cochran}, {K{\"u}rster}, {Paulson},
  {Wittenmyer}, {MacQueen}, \& {Tull}}]{endl2006}
{Endl}, M., {Cochran}, W.~D., {K{\"u}rster}, M., {et~al.} 2006,
  \href{http://dx.doi.org/10.1086/506465}{\JournalTitle{\apj}, 649, 436}

\bibitem[{{Endl} {et~al.}(2000){Endl}, {K{\"u}rster}, \& {Els}}]{endl2000}
{Endl}, M., {K{\"u}rster}, M., \& {Els}, S. 2000, \JournalTitle{\aap}, 362, 585

\bibitem[{{Engle} \& {Guinan}(2018)}]{engle2018}
{Engle}, S.~G., \& {Guinan}, E.~F. 2018,
  \href{http://dx.doi.org/10.3847/2515-5172/aab1f8}{\JournalTitle{Research
  Notes of the American Astronomical Society}, 2, 34}

\bibitem[{{Ford}(2006)}]{ford2006}
{Ford}, E.~B. 2006,
  \href{http://dx.doi.org/10.1086/500802}{\JournalTitle{\apj}, 642, 505}

\bibitem[{Foreman-Mackey(2016)}]{dfm2016}
Foreman-Mackey, D. 2016,
  \href{http://dx.doi.org/10.21105/joss.00024}{\JournalTitle{The Journal of
  Open Source Software}, 24}

\bibitem[{{Foreman-Mackey} {et~al.}(2017){Foreman-Mackey}, {Agol},
  {Ambikasaran}, \& {Angus}}]{dfm2017}
{Foreman-Mackey}, D., {Agol}, E., {Ambikasaran}, S., \& {Angus}, R. 2017,
  \href{http://dx.doi.org/10.3847/1538-3881/aa9332}{\JournalTitle{\aj}, 154,
  220}

\bibitem[{{Foreman-Mackey} {et~al.}(2013){Foreman-Mackey}, {Hogg}, {Lang}, \&
  {Goodman}}]{dfm2013}
{Foreman-Mackey}, D., {Hogg}, D.~W., {Lang}, D., \& {Goodman}, J. 2013,
  \href{http://dx.doi.org/10.1086/670067}{\JournalTitle{PASP}, 125, 306}

\bibitem[{{Fulton} {et~al.}(2018){Fulton}, {Petigura}, {Blunt}, \&
  {Sinukoff}}]{fulton2018}
{Fulton}, B.~J., {Petigura}, E.~A., {Blunt}, S., \& {Sinukoff}, E. 2018,
  \href{http://dx.doi.org/10.1088/1538-3873/aaaaa8}{\JournalTitle{\pasp}, 130,
  044504}

\bibitem[{{Gaia Collaboration}(2018)}]{gaiaDR2}
{Gaia Collaboration}. 2018, \JournalTitle{VizieR Online Data Catalog}, 1345

\bibitem[{Ginsburg {et~al.}(2016)Ginsburg, Parikh, Woillez, Groener, Liedtke,
  Sipocz, Robitaille, Deil, Svoboda, Tollerud, Persson, adamginsburg,
  Séguin-Charbonneau, Armstrong, Mirocha, Droettboom, james allen, Moolekamp,
  Egeland, Singer, Barbary, Grollier, Shiga, Günther, Parejko, Booker, Rol,
  Edward, Miller, \& Willett}]{ginsburg2016}
Ginsburg, A., Parikh, M., Woillez, J., {et~al.} 2016, astroquery v0.3.1

\bibitem[{{Gully-Santiago} {et~al.}(2017){Gully-Santiago}, {Herczeg},
  {Czekala}, {Somers}, {Grankin}, {Covey}, {Donati}, {Alencar}, {Hussain},
  {Shappee}, {Mace}, {Lee}, {Holoien}, {Jose}, \& {Liu}}]{gully2017}
{Gully-Santiago}, M.~A., {Herczeg}, G.~J., {Czekala}, I., {et~al.} 2017,
  \href{http://dx.doi.org/10.3847/1538-4357/836/2/200}{\JournalTitle{\apj},
  836, 200}

\bibitem[{{Harrington}(1990)}]{harrington1990}
{Harrington}, R.~S. 1990,
  \href{http://dx.doi.org/10.1086/115538}{\JournalTitle{\aj}, 100, 559}

\bibitem[{{Hawley} {et~al.}(2003){Hawley}, {Allred}, {Johns-Krull}, {Fisher},
  {Abbett}, {Alekseev}, {Avgoloupis}, {Deustua}, {Gunn}, {Seiradakis}, {Sirk},
  \& {Valenti}}]{hawley2003}
{Hawley}, S.~L., {Allred}, J.~C., {Johns-Krull}, C.~M., {et~al.} 2003,
  \href{http://dx.doi.org/10.1086/378351}{\JournalTitle{\apj}, 597, 535}

\bibitem[{{Haywood} {et~al.}(2014){Haywood}, {Collier Cameron}, {Queloz},
  {Barros}, {Deleuil}, {Fares}, {Gillon}, {Lanza}, {Lovis}, {Moutou}, {Pepe},
  {Pollacco}, {Santerne}, {S{\'e}gransan}, \& {Unruh}}]{haywood2014}
{Haywood}, R.~D., {Collier Cameron}, A., {Queloz}, D., {et~al.} 2014,
  \href{http://dx.doi.org/10.1093/mnras/stu1320}{\JournalTitle{\mnras}, 443,
  2517}

\bibitem[{{Honeycutt}(1992)}]{honeycutt1992}
{Honeycutt}, R.~K. 1992,
  \href{http://dx.doi.org/10.1086/133015}{\JournalTitle{\pasp}, 104, 435}

\bibitem[{{Houdebine} {et~al.}(2016){Houdebine}, {Mullan}, {Paletou}, \&
  {Gebran}}]{houdebine2016}
{Houdebine}, E.~R., {Mullan}, D.~J., {Paletou}, F., \& {Gebran}, M. 2016,
  \href{http://dx.doi.org/10.3847/0004-637X/822/2/97}{\JournalTitle{\apj}, 822,
  97}

\bibitem[{{Hsu} {et~al.}(2020){Hsu}, {Ford}, \& {Terrien}}]{hsu2020}
{Hsu}, D.~C., {Ford}, E.~B., \& {Terrien}, R. 2020, \JournalTitle{arXiv
  e-prints}, arXiv:2002.02573

\bibitem[{{Hunt-Walker} {et~al.}(2012){Hunt-Walker}, {Hilton}, {Kowalski},
  {Hawley}, \& {Matthews}}]{huntwalker2012}
{Hunt-Walker}, N.~M., {Hilton}, E.~J., {Kowalski}, A.~F., {Hawley}, S.~L., \&
  {Matthews}, J.~M. 2012,
  \href{http://dx.doi.org/10.1086/666495}{\JournalTitle{\pasp}, 124, 545}

\bibitem[{{Hunter}(2007)}]{hunter2007}
{Hunter}, J.~D. 2007,
  \href{http://dx.doi.org/10.1109/MCSE.2007.55}{\JournalTitle{Computing in
  Science and Engineering}, 9, 90}

\bibitem[{{Johns-Krull} {et~al.}(2016){Johns-Krull}, {McLane}, {Prato},
  {Crockett}, {Jaffe}, {Hartigan}, {Beichman}, {Mahmud}, {Chen}, {Skiff},
  {Cauley}, {Jones}, \& {Mace}}]{johnskrull2016}
{Johns-Krull}, C.~M., {McLane}, J.~N., {Prato}, L., {et~al.} 2016,
  \href{http://dx.doi.org/10.3847/0004-637X/826/2/206}{\JournalTitle{\apj},
  826, 206}

\bibitem[{{Johnson} {et~al.}(2016){Johnson}, {Endl}, {Cochran}, {Meschiari},
  {Robertson}, {MacQueen}, {Brugamyer}, {Caldwell}, {Hatzes}, {Ram{\'\i}rez},
  \& {Wittenmyer}}]{johnson2016}
{Johnson}, M.~C., {Endl}, M., {Cochran}, W.~D., {et~al.} 2016,
  \href{http://dx.doi.org/10.3847/0004-637X/821/2/74}{\JournalTitle{\apj}, 821,
  74}

\bibitem[{{Kaplan} {et~al.}(2018){Kaplan}, {Bender}, {Terrien}, {Ninan}, {Roy},
  \& {Mahadevan}}]{kaplan2018}
{Kaplan}, K.~F., {Bender}, C.~F., {Terrien}, R., {et~al.} 2018, in The 28th
  International Astronomical Data Analysis Software \& Systems

\bibitem[{Kluyver {et~al.}(2016)Kluyver, Ragan-Kelley, P{\'e}rez, Granger,
  Bussonnier, Frederic, Kelley, Hamrick, Grout, Corlay, Ivanov, Avila, Abdalla,
  Willing, \& development team}]{jupyter2016}
Kluyver, T., Ragan-Kelley, B., P{\'e}rez, F., {et~al.} 2016,
  \href{https://eprints.soton.ac.uk/403913/}{in Positioning and Power in
  Academic Publishing: Players, Agents and Agendas, ed. F.~Loizides \&
  B.~Scmidt} (IOS Press), 87

\bibitem[{{Koen}(2015)}]{koen2015}
{Koen}, C. 2015,
  \href{http://dx.doi.org/10.1093/mnras/stv906}{\JournalTitle{\mnras}, 450,
  3991}

\bibitem[{{Kopparapu} {et~al.}(2013){Kopparapu}, {Ramirez}, {Kasting}, {Eymet},
  {Robinson}, {Mahadevan}, {Terrien}, {Domagal-Goldman}, {Meadows}, \&
  {Deshpande}}]{kopparapu2013}
{Kopparapu}, R.~K., {Ramirez}, R., {Kasting}, J.~F., {et~al.} 2013,
  \href{http://dx.doi.org/10.1088/0004-637X/765/2/131}{\JournalTitle{\apj},
  765, 131}

\bibitem[{{Kreidberg} {et~al.}(2014){Kreidberg}, {Bean}, {D{\'e}sert},
  {Benneke}, {Deming}, {Stevenson}, {Seager}, {Berta-Thompson}, {Seifahrt}, \&
  {Homeier}}]{kreidberg2014}
{Kreidberg}, L., {Bean}, J.~L., {D{\'e}sert}, J.-M., {et~al.} 2014,
  \href{http://dx.doi.org/10.1038/nature12888}{\JournalTitle{\nat}, 505, 69}

\bibitem[{{K{\"u}ker} \& {R{\"u}diger}(2005)}]{kuker2005}
{K{\"u}ker}, M., \& {R{\"u}diger}, G. 2005,
  \href{http://dx.doi.org/10.1002/asna.200410387}{\JournalTitle{Astronomische
  Nachrichten}, 326, 265}

\bibitem[{{Laughlin} {et~al.}(1997){Laughlin}, {Bodenheimer}, \&
  {Adams}}]{laughlin1997}
{Laughlin}, G., {Bodenheimer}, P., \& {Adams}, F.~C. 1997,
  \href{http://dx.doi.org/10.1086/304125}{\JournalTitle{\apj}, 482, 420}

\bibitem[{{Lavail} {et~al.}(2018){Lavail}, {Kochukhov}, \& {Wade}}]{lavail2018}
{Lavail}, A., {Kochukhov}, O., \& {Wade}, G.~A. 2018,
  \href{http://dx.doi.org/10.1093/mnras/sty1825}{\JournalTitle{\mnras}, 479,
  4836}

\bibitem[{{Lightkurve Collaboration} {et~al.}(2018){Lightkurve Collaboration},
  {Cardoso}, {Hedges}, {Gully-Santiago}, {Saunders}, {Cody}, {Barclay}, {Hall},
  {Sagear}, {Turtelboom}, {Zhang}, {Tzanidakis}, {Mighell}, {Coughlin}, {Bell},
  {Berta-Thompson}, {Williams}, {Dotson}, \& {Barentsen}}]{lightkurve2018}
{Lightkurve Collaboration}, {Cardoso}, J.~V.~d.~M., {Hedges}, C., {et~al.}
  2018, {Lightkurve: Kepler and TESS time series analysis in Python},
  Astrophysics Source Code Library,
  \href{http://arxiv.org/abs/1812.013}{{\sffamily ascl:1812.013}}

\bibitem[{{L{\'o}pez-Morales} {et~al.}(2016){L{\'o}pez-Morales}, {Haywood},
  {Coughlin}, {Zeng}, {Buchhave}, {Giles}, {Affer}, {Bonomo}, {Charbonneau},
  {Collier Cameron}, {Consentino}, {Dressing}, {Dumusque}, {Figueira},
  {Fiorenzano}, {Harutyunyan}, {Johnson}, {Latham}, {Lopez}, {Lovis},
  {Malavolta}, {Mayor}, {Micela}, {Molinari}, {Mortier}, {Motalebi},
  {Nascimbeni}, {Pepe}, {Phillips}, {Piotto}, {Pollacco}, {Queloz}, {Rice},
  {Sasselov}, {Segransan}, {Sozzetti}, {Udry}, {Vanderburg}, \&
  {Watson}}]{lopezmorales2016}
{L{\'o}pez-Morales}, M., {Haywood}, R.~D., {Coughlin}, J.~L., {et~al.} 2016,
  \href{http://dx.doi.org/10.3847/0004-6256/152/6/204}{\JournalTitle{\aj}, 152,
  204}

\bibitem[{{Lurie} {et~al.}(2015){Lurie}, {Davenport}, {Hawley}, {Wilkinson},
  {Wisniewski}, {Kowalski}, \& {Hebb}}]{lurie2015}
{Lurie}, J.~C., {Davenport}, J. R.~A., {Hawley}, S.~L., {et~al.} 2015,
  \href{http://dx.doi.org/10.1088/0004-637X/800/2/95}{\JournalTitle{\apj}, 800,
  95}

\bibitem[{{Mahadevan} {et~al.}(2012){Mahadevan}, {Ramsey}, {Bender}, {Terrien},
  {Wright}, {Halverson}, {Hearty}, {Nelson}, {Burton}, {Redman}, {Osterman},
  {Diddams}, {Kasting}, {Endl}, \& {Deshpande}}]{mahadevan2012}
{Mahadevan}, S., {Ramsey}, L., {Bender}, C., {et~al.} 2012,
  \href{http://dx.doi.org/10.1117/12.926102}{in \procspie, Vol. 8446,
  Ground-based and Airborne Instrumentation for Astronomy IV}, 84461S

\bibitem[{{Mahadevan} {et~al.}(2014){Mahadevan}, {Ramsey}, {Terrien},
  {Halverson}, {Roy}, {Hearty}, {Levi}, {Stefansson}, {Robertson}, {Bender},
  {Schwab}, \& {Nelson}}]{mahadevan2014}
{Mahadevan}, S., {Ramsey}, L.~W., {Terrien}, R., {et~al.} 2014, Society of
  Photo-Optical Instrumentation Engineers (SPIE) Conference Series, Vol. 9147,
  {The Habitable-zone Planet Finder: A status update on the development of a
  stabilized fiber-fed near-infrared spectrograph for the for the Hobby-Eberly
  telescope}, 91471G

\bibitem[{{Mann} {et~al.}(2015){Mann}, {Feiden}, {Gaidos}, {Boyajian}, \& {von
  Braun}}]{mann2015}
{Mann}, A.~W., {Feiden}, G.~A., {Gaidos}, E., {Boyajian}, T., \& {von Braun},
  K. 2015,
  \href{http://dx.doi.org/10.1088/0004-637X/804/1/64}{\JournalTitle{\apj}, 804,
  64}

\bibitem[{{Marchwinski} {et~al.}(2015){Marchwinski}, {Mahadevan}, {Robertson},
  {Ramsey}, \& {Harder}}]{marchwinski2015}
{Marchwinski}, R.~C., {Mahadevan}, S., {Robertson}, P., {Ramsey}, L., \&
  {Harder}, J. 2015,
  \href{http://dx.doi.org/10.1088/0004-637X/798/1/63}{\JournalTitle{\apj}, 798,
  63}

\bibitem[{{Martinez Pillet} {et~al.}(1993){Martinez Pillet}, {Moreno-Insertis},
  \& {Vazquez}}]{martinezpillet1993}
{Martinez Pillet}, V., {Moreno-Insertis}, F., \& {Vazquez}, M. 1993,
  \JournalTitle{\aap}, 274, 521

\bibitem[{{Masuda} \& {Winn}(2020)}]{masuda2020}
{Masuda}, K., \& {Winn}, J.~N. 2020,
  \href{http://dx.doi.org/10.3847/1538-3881/ab65be}{\JournalTitle{\aj}, 159,
  81}

\bibitem[{McKinney(2010)}]{pandas2010}
McKinney, W. 2010, in Proceedings of the 9th Python in Science Conference, ed.
  S.~van~der Walt \& J.~Millman, 51

\bibitem[{{McQuillan} {et~al.}(2014){McQuillan}, {Mazeh}, \&
  {Aigrain}}]{mcquillan2014}
{McQuillan}, A., {Mazeh}, T., \& {Aigrain}, S. 2014,
  \href{http://dx.doi.org/10.1088/0067-0049/211/2/24}{\JournalTitle{\apjs},
  211, 24}

\bibitem[{{Metcalf} {et~al.}(2019){Metcalf}, {Anderson}, {Bender}, {Blakeslee},
  {Brand}, {Carlson}, {Cochran}, {Diddams}, {Endl}, {Fredrick}, {Halverson},
  {Hickstein}, {Hearty}, {Jennings}, {Kanodia}, {Kaplan}, {Levi}, {Lubar},
  {Mahadevan}, {Monson}, {Ninan}, {Nitroy}, {Osterman}, {Papp}, {Quinlan},
  {Ramsey}, {Robertson}, {Roy}, {Schwab}, {Sigurdsson}, {Srinivasan},
  {Stefansson}, {Sterner}, {Terrien}, {Wolszczan}, {Wright}, \&
  {Ycas}}]{metcalf2019}
{Metcalf}, A.~J., {Anderson}, T., {Bender}, C.~F., {et~al.} 2019,
  \href{http://dx.doi.org/10.1364/OPTICA.6.000233}{\JournalTitle{Optica}, 6,
  233}

\bibitem[{{Morin} {et~al.}(2008){Morin}, {Donati}, {Petit}, {Delfosse},
  {Forveille}, {Albert}, {Auri{\`e}re}, {Cabanac}, {Dintrans}, {Fares},
  {Gastine}, {Jardine}, {Ligni{\`e}res}, {Paletou}, {Ramirez Velez}, \&
  {Th{\'e}ado}}]{morin2008}
{Morin}, J., {Donati}, J.-F., {Petit}, P., {et~al.} 2008,
  \href{http://dx.doi.org/10.1111/j.1365-2966.2008.13809.x}{\JournalTitle{\mnras},
  390, 567}

\bibitem[{{Morton}(2015)}]{morton2015}
{Morton}, T.~D. 2015, {isochrones: Stellar model grid package}

\bibitem[{{Muirhead} {et~al.}(2018){Muirhead}, {Dressing}, {Mann},
  {Rojas-Ayala}, {L{\'e}pine}, {Paegert}, {De Lee}, \&
  {Oelkers}}]{muirhead2018}
{Muirhead}, P.~S., {Dressing}, C.~D., {Mann}, A.~W., {et~al.} 2018,
  \href{http://dx.doi.org/10.3847/1538-3881/aab710}{\JournalTitle{\aj}, 155,
  180}

\bibitem[{{Newton} {et~al.}(2017){Newton}, {Irwin}, {Charbonneau}, {Berlind},
  {Calkins}, \& {Mink}}]{newton2017}
{Newton}, E.~R., {Irwin}, J., {Charbonneau}, D., {et~al.} 2017,
  \href{http://dx.doi.org/10.3847/1538-4357/834/1/85}{\JournalTitle{\apj}, 834,
  85}

\bibitem[{{Newton} {et~al.}(2016){Newton}, {Irwin}, {Charbonneau},
  {Berta-Thompson}, \& {Dittmann}}]{newton2016}
{Newton}, E.~R., {Irwin}, J., {Charbonneau}, D., {Berta-Thompson}, Z.~K., \&
  {Dittmann}, J.~A. 2016,
  \href{http://dx.doi.org/10.3847/2041-8205/821/1/L19}{\JournalTitle{\apjl},
  821, L19}

\bibitem[{{Ninan} {et~al.}(2018){Ninan}, {Bender}, {Mahadevan}, {Ford},
  {Monson}, {Kaplan}, {Terrien}, {Roy}, {Robertson}, {Kanodia}, \&
  {Stefansson}}]{ninan2018}
{Ninan}, J.~P., {Bender}, C.~F., {Mahadevan}, S., {et~al.} 2018,
  \href{http://dx.doi.org/10.1117/12.2312787}{in Society of Photo-Optical
  Instrumentation Engineers (SPIE) Conference Series, Vol. 10709, High Energy,
  Optical, and Infrared Detectors for Astronomy VIII}, 107092U

\bibitem[{{Queloz} {et~al.}(2001){Queloz}, {Henry}, {Sivan}, {Baliunas},
  {Beuzit}, {Donahue}, {Mayor}, {Naef}, {Perrier}, \& {Udry}}]{queloz2001}
{Queloz}, D., {Henry}, G.~W., {Sivan}, J.~P., {et~al.} 2001,
  \href{http://dx.doi.org/10.1051/0004-6361:20011308}{\JournalTitle{\aap}, 379,
  279}

\bibitem[{{Reiners}(2009)}]{reiners2009}
{Reiners}, A. 2009,
  \href{http://dx.doi.org/10.1051/0004-6361/200810257}{\JournalTitle{\aap},
  498, 853}

\bibitem[{{Reiners} {et~al.}(2010){Reiners}, {Bean}, {Huber}, {Dreizler},
  {Seifahrt}, \& {Czesla}}]{reiners2010}
{Reiners}, A., {Bean}, J.~L., {Huber}, K.~F., {et~al.} 2010,
  \href{http://dx.doi.org/10.1088/0004-637X/710/1/432}{\JournalTitle{\apj},
  710, 432}

\bibitem[{{Reiners} {et~al.}(2013){Reiners}, {Shulyak}, {Anglada-Escud{\'e}},
  {Jeffers}, {Morin}, {Zechmeister}, {Kochukhov}, \& {Piskunov}}]{reiners2013}
{Reiners}, A., {Shulyak}, D., {Anglada-Escud{\'e}}, G., {et~al.} 2013,
  \href{http://dx.doi.org/10.1051/0004-6361/201220437}{\JournalTitle{\aap},
  552, A103}

\bibitem[{{Reiners} {et~al.}(2018){Reiners}, {Zechmeister}, {Caballero},
  {Ribas}, {Morales}, {Jeffers}, {Schofer}, {Tal-Or}, {Quirrenbach}, {Amado},
  {Kaminski}, {Seifert}, {Abril}, {Aceituno}, {Alonso-Floriano}, {Ammler-von
  Eiff}, {Antona}, {Anglada-Escude}, {Anwand-Heerwart}, {Arroyo-Torres},
  {Azzaro}, {Baroch}, {Barrado}, {Bauer}, {Becerril}, {Bejar}, {Benitez},
  {Berdinas}, {Bergond}, {Blumcke}, {Brinkmoller}, {Del Burgo}, {Cano},
  {Cardenas Vazquez}, {Casal}, {Cifuentes}, {Claret}, {Colome},
  {Cortes-Contreras}, {Czesla}, {Diez-Alonso}, {Dreizler}, {Feiz}, {Fernandez},
  {Ferro}, {Fuhrmeister}, {Galadi-Enriquez}, {Garcia-Piquer}, {Garcia Vargas},
  {Gesa}, {Gomez Galera}, {Gonzalez Hernandez}, {Gonzalez-Peinado},
  {Grozinger}, {Grohnert}, {Guardia}, {Guenther}, {Guijarro}, {de Guindos},
  {Gutierrez-Soto}, {Hagen}, {Hatzes}, {Hauschildt}, {Hedrosa}, {Helmling},
  {Henning}, {Hermelo}, {Hernandez Arabi}, {Hernand ez Castano}, {Hernandez
  Hernando}, {Herrero}, {Huber}, {Huke}, {Johnson}, {de Juan}, {Kim}, {Klein},
  {Kluter}, {Klutsch}, {Kurster}, {Lafarga}, {Lamert}, {Lampon}, {Lara},
  {Laun}, {Lemke}, {Lenzen}, {Launhardt}, {Lopez Del Fresno}, {Lopez-Gonzalez},
  {Lopez-Puertas}, {Lopez Salas}, {Lopez-Santiago}, {Luque}, {Magan
  Madinabeitia}, {Mall}, {Mancini}, {Mandel}, {Marfil}, {Marin Molina}, {Maroto
  Fernandez}, {Martin}, {Martin-Ruiz}, {Marvin}, {Mathar}, {Mirabet}, {Montes},
  {Moreno-Raya}, {Moya}, {Mundt}, {Nagel}, {Naranjo}, {Nortmann}, {Nowak},
  {Ofir}, {Oreiro}, {Palle}, {Pand uro}, {Pascual}, {Passegger}, {Pavlov},
  {Pedraz}, {Perez-Calpena}, {Perez Medialdea}, {Perger}, {Perryman}, {Pluto},
  {Rabaza}, {Ramon}, {Rebolo}, {Redondo}, {Reffert}, {Reinhart}, {Rhode},
  {Rix}, {Rodler}, {Rodriguez}, {Rodriguez-Lopez}, {Rodriguez Trinidad},
  {Rohloff}, {Rosich}, {Sadegi}, {Sanchez-Blanco}, {Sanchez Carrasco},
  {Sanchez-Lopez}, {Sanz-Forcada}, {Sarkis}, {Sarmiento}, {Schafer}, {Schmitt},
  {Schiller}, {Schweitzer}, {Solano}, {Stahl}, {Strachan}, {Sturmer}, {Suarez},
  {Tabernero}, {Tala}, {Trifonov}, {Tulloch}, {Ulbrich}, {Veredas}, {Vico
  Linares}, {Vilardell}, {Wagner}, {Winkler}, {Wolthoff}, {Xu}, {Yan}, \&
  {Zapatero Osorio}}]{reiners2018}
{Reiners}, A., {Zechmeister}, M., {Caballero}, J.~A., {et~al.} 2018,
  \JournalTitle{VizieR Online Data Catalog}, J/A+A/612/A49

\bibitem[{{Ribas} {et~al.}(2018){Ribas}, {Tuomi}, {Reiners}, {Butler},
  {Morales}, {Perger}, {Dreizler}, {Rodr{\'\i}guez-L{\'o}pez}, {Gonz{\'a}lez
  Hern{\'a}ndez}, {Rosich}, {Feng}, {Trifonov}, {Vogt}, {Caballero}, {Hatzes},
  {Herrero}, {Jeffers}, {Lafarga}, {Murgas}, {Nelson}, {Rodr{\'\i}guez},
  {Strachan}, {Tal-Or}, {Teske}, {Toledo-Padr{\'o}n}, {Zechmeister},
  {Quirrenbach}, {Amado}, {Azzaro}, {B{\'e}jar}, {Barnes}, {Berdi{\~n}as},
  {Burt}, {Coleman}, {Cort{\'e}s-Contreras}, {Crane}, {Engle}, {Guinan},
  {Haswell}, {Henning}, {Holden}, {Jenkins}, {Jones}, {Kaminski}, {Kiraga},
  {K{\"u}rster}, {Lee}, {L{\'o}pez-Gonz{\'a}lez}, {Montes}, {Morin}, {Ofir},
  {Pall{\'e}}, {Rebolo}, {Reffert}, {Schweitzer}, {Seifert}, {Shectman},
  {Staab}, {Street}, {Su{\'a}rez Mascare{\~n}o}, {Tsapras}, {Wang}, \&
  {Anglada-Escud{\'e}}}]{ribas2018}
{Ribas}, I., {Tuomi}, M., {Reiners}, A., {et~al.} 2018,
  \href{http://dx.doi.org/10.1038/s41586-018-0677-y}{\JournalTitle{\nat}, 563,
  365}

\bibitem[{{Ricker} {et~al.}(2015){Ricker}, {Winn}, {Vanderspek}, {Latham},
  {Bakos}, {Bean}, {Berta-Thompson}, {Brown}, {Buchhave}, {Butler}, {Butler},
  {Chaplin}, {Charbonneau}, {Christensen-Dalsgaard}, {Clampin}, {Deming},
  {Doty}, {De Lee}, {Dressing}, {Dunham}, {Endl}, {Fressin}, {Ge}, {Henning},
  {Holman}, {Howard}, {Ida}, {Jenkins}, {Jernigan}, {Johnson}, {Kaltenegger},
  {Kawai}, {Kjeldsen}, {Laughlin}, {Levine}, {Lin}, {Lissauer}, {MacQueen},
  {Marcy}, {McCullough}, {Morton}, {Narita}, {Paegert}, {Palle}, {Pepe},
  {Pepper}, {Quirrenbach}, {Rinehart}, {Sasselov}, {Sato}, {Seager},
  {Sozzetti}, {Stassun}, {Sullivan}, {Szentgyorgyi}, {Torres}, {Udry}, \&
  {Villasenor}}]{ricker2015}
{Ricker}, G.~R., {Winn}, J.~N., {Vanderspek}, R., {et~al.} 2015,
  \href{http://dx.doi.org/10.1117/1.JATIS.1.1.014003}{\JournalTitle{Journal of
  Astronomical Telescopes, Instruments, and Systems}, 1, 014003}

\bibitem[{{Robertson} {et~al.}(2014){Robertson}, {Mahadevan}, {Endl}, \&
  {Roy}}]{robertson2014}
{Robertson}, P., {Mahadevan}, S., {Endl}, M., \& {Roy}, A. 2014,
  \href{http://dx.doi.org/10.1126/science.1253253}{\JournalTitle{Science}, 345,
  440}

\bibitem[{{Robertson} {et~al.}(2015){Robertson}, {Roy}, \&
  {Mahadevan}}]{robertson2015}
{Robertson}, P., {Roy}, A., \& {Mahadevan}, S. 2015,
  \href{http://dx.doi.org/10.1088/2041-8205/805/2/L22}{\JournalTitle{\apjl},
  805, L22}

\bibitem[{{Santos} {et~al.}(2014){Santos}, {Mortier}, {Faria}, {Dumusque},
  {Adibekyan}, {Delgado-Mena}, {Figueira}, {Benamati}, {Boisse}, {Cunha},
  {Gomes da Silva}, {Lo Curto}, {Lovis}, {Martins}, {Mayor}, {Melo}, {Oshagh},
  {Pepe}, {Queloz}, {Santerne}, {S{\'e}gransan}, {Sozzetti}, {Sousa}, \&
  {Udry}}]{santos2014}
{Santos}, N.~C., {Mortier}, A., {Faria}, J.~P., {et~al.} 2014,
  \href{http://dx.doi.org/10.1051/0004-6361/201423808}{\JournalTitle{\aap},
  566, A35}

\bibitem[{{Schweitzer} {et~al.}(2019){Schweitzer}, {Passegger}, {Cifuentes},
  {B{\'e}jar}, {Cort{\'e}s-Contreras}, {Caballero}, {del Burgo}, {Czesla},
  {K{\"u}rster}, {Montes}, {Zapatero Osorio}, {Ribas}, {Reiners},
  {Quirrenbach}, {Amado}, {Aceituno}, {Anglada-Escud{\'e}}, {Bauer},
  {Dreizler}, {Jeffers}, {Guenther}, {Henning}, {Kaminski}, {Lafarga},
  {Marfil}, {Morales}, {Schmitt}, {Seifert}, {Solano}, {Tabernero}, \&
  {Zechmeister}}]{schweitzer2019}
{Schweitzer}, A., {Passegger}, V.~M., {Cifuentes}, C., {et~al.} 2019,
  \href{http://dx.doi.org/10.1051/0004-6361/201834965}{\JournalTitle{\aap},
  625, A68}

\bibitem[{{Shetrone} {et~al.}(2007){Shetrone}, {Cornell}, {Fowler}, {Gaffney},
  {Laws}, {Mader}, {Mason}, {Odewahn}, {Roman}, {Rostopchin}, {Schneider},
  {Umbarger}, \& {Westfall}}]{shetrone2007}
{Shetrone}, M., {Cornell}, M.~E., {Fowler}, J.~R., {et~al.} 2007,
  \href{http://dx.doi.org/10.1086/519291}{\JournalTitle{\pasp}, 119, 556}

\bibitem[{{Shulyak} {et~al.}(2017){Shulyak}, {Reiners}, {Engeln}, {Malo},
  {Yadav}, {Morin}, \& {Kochukhov}}]{shulyak2017}
{Shulyak}, D., {Reiners}, A., {Engeln}, A., {et~al.} 2017,
  \href{http://dx.doi.org/10.1038/s41550-017-0184}{\JournalTitle{Nature
  Astronomy}, 1, 0184}

\bibitem[{{Stefansson} {et~al.}(2016){Stefansson}, {Hearty}, {Robertson},
  {Mahadevan}, {Anderson}, {Levi}, {Bender}, {Nelson}, {Monson}, {Blank},
  {Halverson}, {Henderson}, {Ramsey}, {Roy}, {Schwab}, \&
  {Terrien}}]{stefansson2016}
{Stefansson}, G., {Hearty}, F., {Robertson}, P., {et~al.} 2016,
  \href{http://dx.doi.org/10.3847/1538-4357/833/2/175}{\JournalTitle{\apj},
  833, 175}

\bibitem[{{Stefansson} {et~al.}(2020){Stefansson}, {Ca{\~n}as}, {Wisniewski},
  {Robertson}, {Mahadevan}, {Maney}, {Kanodia}, {Beard}, {Bender}, {Brunt},
  {Clemens}, {Cochran}, {Diddams}, {Endl}, {Ford}, {Fredrick}, {Halverson},
  {Hearty}, {Hebb}, {Huehnerhoff}, {Jennings}, {Kaplan}, {Levi}, {Lubar},
  {Metcalf}, {Monson}, {Morris}, {Ninan}, {Nitroy}, {Ramsey}, {Roy}, {Schwab},
  {Sigurdsson}, {Terrien}, \& {Wright}}]{stefansson2020}
{Stefansson}, G., {Ca{\~n}as}, C., {Wisniewski}, J., {et~al.} 2020,
  \href{http://dx.doi.org/10.3847/1538-3881/ab5f15}{\JournalTitle{\aj}, 159,
  100}

\bibitem[{{Su{\'a}rez Mascare{\~n}o} {et~al.}(2018){Su{\'a}rez Mascare{\~n}o},
  {Rebolo}, {Gonz{\'a}lez Hern{\'a}ndez}, {Toledo-Padr{\'o}n}, {Perger},
  {Ribas}, {Affer}, {Micela}, {Damasso}, {Maldonado}, {Gonz{\'a}lez-Alvarez},
  {Leto}, {Pagano}, {Scandariato}, {Sozzetti}, {Lanza}, {Malavolta}, {Claudi},
  {Cosentino}, {Desidera}, {Giacobbe}, {Maggio}, {Rainer}, {Esposito},
  {Benatti}, {Pedani}, {Morales}, {Herrero}, {Lafarga}, {Rosich}, \&
  {Pinamonti}}]{suarez2018}
{Su{\'a}rez Mascare{\~n}o}, A., {Rebolo}, R., {Gonz{\'a}lez Hern{\'a}ndez},
  J.~I., {et~al.} 2018,
  \href{http://dx.doi.org/10.1051/0004-6361/201732143}{\JournalTitle{\aap},
  612, A89}

\bibitem[{{Tal-Or} {et~al.}(2019){Tal-Or}, {Trifonov}, {Zucker}, {Mazeh}, \&
  {Zechmeister}}]{talor2019}
{Tal-Or}, L., {Trifonov}, T., {Zucker}, S., {Mazeh}, T., \& {Zechmeister}, M.
  2019, \href{http://dx.doi.org/10.1093/mnrasl/sly227}{\JournalTitle{\mnras},
  484, L8}

\bibitem[{{Tal-Or} {et~al.}(2018){Tal-Or}, {Zechmeister}, {Reiners}, {Jeffers},
  {Sch{\"o}fer}, {Quirrenbach}, {Amado}, {Ribas}, {Caballero}, {Aceituno},
  {Bauer}, {B{\'e}jar}, {Czesla}, {Dreizler}, {Fuhrmeister}, {Hatzes},
  {Johnson}, {K{\"u}rster}, {Lafarga}, {Montes}, {Morales}, {Reffert},
  {Sadegi}, {Seifert}, \& {Shulyak}}]{talor2018}
{Tal-Or}, L., {Zechmeister}, M., {Reiners}, A., {et~al.} 2018,
  \href{http://dx.doi.org/10.1051/0004-6361/201732362}{\JournalTitle{\aap},
  614, A122}

\bibitem[{{Tanimoto} {et~al.}(2020){Tanimoto}, {Yamashita}, {Ui}, {Uchiyama},
  {Kawabata}, {Mori}, {Nakaoka}, {Abe}, {Itoh}, {Kand a}, {Kawaguchi},
  {Kawahara}, {Otsubo}, {Shiki}, {Takagi}, {Takaki}, {Akitaya}, {Yamanaka}, \&
  {Kawabata}}]{tanimoto2020}
{Tanimoto}, Y., {Yamashita}, T., {Ui}, T., {et~al.} 2020,
  \href{http://dx.doi.org/10.1093/pasj/psz145}{\JournalTitle{\pasj}},
  \href{http://arxiv.org/abs/2001.00148}{{\sffamily arXiv:2001.00148
  [astro-ph.EP]}}

\bibitem[{{Thompson} {et~al.}(2003){Thompson}, {Christensen-Dalsgaard},
  {Miesch}, \& {Toomre}}]{thompson2003}
{Thompson}, M.~J., {Christensen-Dalsgaard}, J., {Miesch}, M.~S., \& {Toomre},
  J. 2003,
  \href{http://dx.doi.org/10.1146/annurev.astro.41.011802.094848}{\JournalTitle{\araa},
  41, 599}

\bibitem[{{Trifonov} {et~al.}(2020){Trifonov}, {Tal-Or}, {Zechmeister},
  {Kaminski}, {Zucker}, \& {Mazeh}}]{trifonov2020}
{Trifonov}, T., {Tal-Or}, L., {Zechmeister}, M., {et~al.} 2020,
  \JournalTitle{arXiv e-prints}, arXiv:2001.05942

\bibitem[{{Tuomi} {et~al.}(2018){Tuomi}, {Jones}, {Barnes},
  {Anglada-Escud{\'e}}, {Butler}, {Kiraga}, \& {Vogt}}]{tuomi2018}
{Tuomi}, M., {Jones}, H.~R.~A., {Barnes}, J.~R., {et~al.} 2018,
  \href{http://dx.doi.org/10.3847/1538-3881/aab09c}{\JournalTitle{\aj}, 155,
  192}

\bibitem[{{Valenti} {et~al.}(1995){Valenti}, {Butler}, \&
  {Marcy}}]{valenti1995}
{Valenti}, J.~A., {Butler}, R.~P., \& {Marcy}, G.~W. 1995,
  \href{http://dx.doi.org/10.1086/133645}{\JournalTitle{\pasp}, 107, 966}

\bibitem[{{van den Besselaar} {et~al.}(2003){van den Besselaar}, {Raassen},
  {Mewe}, {van der Meer}, {G{\"u}del}, \& {Audard}}]{vdb2003}
{van den Besselaar}, E.~J.~M., {Raassen}, A.~J.~J., {Mewe}, R., {et~al.} 2003,
  \href{http://dx.doi.org/10.1051/0004-6361:20031398}{\JournalTitle{\aap}, 411,
  587}

\bibitem[{{Van Der Walt} {et~al.}(2011){Van Der Walt}, {Colbert}, \&
  {Varoquaux}}]{vanderwalt2011}
{Van Der Walt}, S., {Colbert}, S.~C., \& {Varoquaux}, G. 2011,
  \JournalTitle{ArXiv e-prints},
  \href{http://arxiv.org/abs/1102.1523}{{\sffamily arXiv:1102.1523 [cs.MS]}}

\bibitem[{{van Eyken} {et~al.}(2012){van Eyken}, {Ciardi}, {von Braun}, {Kane},
  {Plavchan}, {Bender}, {Brown}, {Crepp}, {Fulton}, {Howard}, {Howell},
  {Mahadevan}, {Marcy}, {Shporer}, {Szkody}, {Akeson}, {Beichman}, {Boden},
  {Gelino}, {Hoard}, {Ram{\'\i}rez}, {Rebull}, {Stauffer}, {Bloom}, {Cenko},
  {Kasliwal}, {Kulkarni}, {Law}, {Nugent}, {Ofek}, {Poznanski}, {Quimby},
  {Walters}, {Grillmair}, {Laher}, {Levitan}, {Sesar}, \&
  {Surace}}]{vaneyken2012}
{van Eyken}, J.~C., {Ciardi}, D.~R., {von Braun}, K., {et~al.} 2012,
  \href{http://dx.doi.org/10.1088/0004-637X/755/1/42}{\JournalTitle{\apj}, 755,
  42}

\bibitem[{{Vanderburg} {et~al.}(2016){Vanderburg}, {Plavchan}, {Johnson},
  {Ciardi}, {Swift}, \& {Kane}}]{vanderburg2016_prot}
{Vanderburg}, A., {Plavchan}, P., {Johnson}, J.~A., {et~al.} 2016,
  \href{http://dx.doi.org/10.1093/mnras/stw863}{\JournalTitle{\mnras}, 459,
  3565}

\bibitem[{{Vogt} {et~al.}(2010){Vogt}, {Butler}, {Rivera}, {Haghighipour},
  {Henry}, \& {Williamson}}]{vogt2010}
{Vogt}, S.~S., {Butler}, R.~P., {Rivera}, E.~J., {et~al.} 2010,
  \href{http://dx.doi.org/10.1088/0004-637X/723/1/954}{\JournalTitle{\apj},
  723, 954}

\bibitem[{{Yu} {et~al.}(2015){Yu}, {Winn}, {Gillon}, {Albrecht}, {Rappaport},
  {Bieryla}, {Dai}, {Delrez}, {Hillenbrand}, {Holman}, {Howard}, {Huang},
  {Isaacson}, {Jehin}, {Lendl}, {Montet}, {Muirhead}, {Sanchis-Ojeda}, \&
  {Triaud}}]{yu2015}
{Yu}, L., {Winn}, J.~N., {Gillon}, M., {et~al.} 2015,
  \href{http://dx.doi.org/10.1088/0004-637X/812/1/48}{\JournalTitle{\apj}, 812,
  48}

\bibitem[{{Zechmeister} \& {K{\"u}rster}(2009)}]{zk2009}
{Zechmeister}, M., \& {K{\"u}rster}, M. 2009,
  \href{http://dx.doi.org/10.1051/0004-6361:200811296}{\JournalTitle{\aap},
  496, 577}

\bibitem[{{Zechmeister} {et~al.}(2018){Zechmeister}, {Reiners}, {Amado},
  {Azzaro}, {Bauer}, {B{\'e}jar}, {Caballero}, {Guenther}, {Hagen}, {Jeffers},
  {Kaminski}, {K{\"u}rster}, {Launhardt}, {Montes}, {Morales}, {Quirrenbach},
  {Reffert}, {Ribas}, {Seifert}, {Tal-Or}, \& {Wolthoff}}]{zechmeister2018}
{Zechmeister}, M., {Reiners}, A., {Amado}, P.~J., {et~al.} 2018,
  \href{http://dx.doi.org/10.1051/0004-6361/201731483}{\JournalTitle{\aap},
  609, A12}

\end{thebibliography}

\newpage

\appendix
\section{Data Tables}

\begin{table}[h!]
    \centering
    \begin{tabular}{| c c c |}
         \hline
         BJD & RV & $\sigma_{\textrm{RM}}$  \\
          & ($\unit{m} \unit{s}^{-1}$) & ($\unit{m} \unit{s}^{-1}$) \\
          \hline
         \emph{Pre-maintenance} & & \\
         2458233.7316037375  &  -12.81 &  1.0  \\
         2458234.7287878820  &   22.40 &  1.1  \\
         2458236.7293598424  &   16.11 &  1.1  \\
         2458237.7299096250  &  -29.77 &  1.4  \\
         2458238.7238860875  &   -1.26 &  1.2  \\
         \emph{Post-maintenance} & & \\
         2458473.8690427130  &  -10.84 &  1.2  \\
         2458476.8599258005  &   10.05 &  1.5  \\
         2458477.8559555200  &   -7.45 &  1.1  \\
         2458481.0492757780  &    7.55 &  1.4  \\
         2458489.0303301950  &   -2.75 &  1.2  \\
         2458500.7923769862  &    7.70 &  1.3  \\
         2458503.7973400960  &    4.74 &  1.2  \\
         2458507.7780896930  &    8.66 &  2.0  \\
         2458508.7752996893  &   -4.10 &  1.3  \\
         2458509.7704128380  &    4.50 &  1.2  \\
         2458512.9689388200  &   -0.29 &  1.6  \\
         2458522.7393540850  &   -3.61 &  1.4  \\
         2458529.7058527880  &    5.52 &  4.8  \\
         2458536.9098783443  &    5.04 &  1.2  \\
         2458538.6966876250  &    4.76 &  1.3  \\
         2458539.8999842554  &   -9.90 &  1.5  \\
         2458541.6804011953  &   -2.11 &  1.4  \\
         2458546.6761750420  &   -5.99 &  2.1  \\
         2458559.6363051767  &   -0.79 &  1.3  \\
         2458590.7565944204  &    2.32 &  1.7  \\
         2458593.7532936020  &   -6.41 &  1.0  \\
         2458596.7309973100  &    8.96 &  1.4  \\
         2458618.6695084795  &    1.81 &  1.1  \\
         2458621.6677907016  &    2.11 &  1.3  \\
         2458622.6635317607  &  -13.06 &  1.2  \\
         2458630.6384578010  &    7.87 &  1.3  \\
         2458887.7390454006  &   -6.55 &  1.4  \\
         2458892.9350831793  &    1.36 &  1.4  \\
         2458902.6923081093  &    6.64 &  1.8  \\
         2458904.6938378840  &   11.80 &  1.4  \\
         \hline

    \end{tabular}
    \caption{HET/HPF radial velocities of AD Leo}
    \label{tab:gj388_rv}
\end{table}

\begin{table}[h!]
    \centering
    \begin{tabular}{| c c c |}
         \hline
         BJD & RV & $\sigma_{\textrm{RM}}$  \\
          & ($\unit{m} \unit{s}^{-1}$) & ($\unit{m} \unit{s}^{-1}$) \\
          \hline
         2458384.7240376705  &    0.01  &  3.0  \\
         2458385.7172474210  &  -26.90  &  2.6  \\
         2458398.6938574360  &   57.74  &  5.5  \\
         2458402.6836300040  &  -42.91  &  2.3  \\
         2458405.6832734635  &   29.44  &  2.5  \\
         2458417.6554489443  &  -19.42  &  2.7  \\
         2458424.6172750550  &  -41.28  &  2.7  \\
         2458427.6106645260  &   -9.24  &  3.3  \\
         2458429.6132536368  &  -48.84  &  4.8  \\
         2458430.6062095350  &   49.35  &  2.4  \\
         2458431.6040842435  &   12.83  &  2.5  \\
         2458433.5950916815  &   -2.86  &  2.9  \\
         2458437.5938322986  &   21.75  &  2.4  \\
         2458439.5759858433  &  -47.96  &  3.1  \\
         2458440.5887918435  &   36.70  &  2.5  \\
         2458441.5632593343  &   -4.23  &  3.8  \\
         2458442.5663821395  &   18.00  &  2.1  \\
         2458444.5741144414  &  -38.90  &  2.7  \\
         2458446.5664232653  &  -11.97  &  2.9  \\
         2458450.5487833503  &   28.67  &  2.4  \\
         2458609.8862807840  &  -23.58  &  5.1  \\
         \hline

    \end{tabular}
    \caption{HET/HPF radial velocities of GJ 1245B}
    \label{tab:gj1245B_rv}
\end{table}

\begin{table}[h!]
    \centering
    \begin{tabular}{| c c c |}
         \hline
         BJD & RV & $\sigma_{\textrm{RM}}$  \\
          & ($\unit{m} \unit{s}^{-1}$) & ($\unit{m} \unit{s}^{-1}$) \\
          \hline
         2458643.82992432  &   100.72  &  16.7  \\
         2458643.83753301  &   116.64  &  19.2  \\
         2458643.84507677  &   119.63  &  18.3  \\
         2458643.85262053  &   158.35  &  19.3  \\
         2458643.86016429  &   215.78  &  18.8  \\
         2458643.86770805  &   228.89  &  18.8  \\
         2458643.87525181  &   248.61  &  17.7  \\
         2458643.88279557  &   261.64  &  18.0  \\
         2458643.89034523  &   163.76  &  24.9  \\
         2458643.89788899  &    95.98  &  23.6  \\
         2458643.90542095  &   125.87  &  19.8  \\
         2458643.91580394  &   162.71  &  18.3  \\
         2458643.92334770  &   209.53  &  18.2  \\
         2458643.93089737  &   155.06  &  17.0  \\
         2458643.93844113  &   121.00  &  18.3  \\
         2458643.94598488  &   101.95  &  16.7  \\
         2458643.95353455  &    42.26  &  18.1  \\
         2458643.96107831  &    11.51  &  19.4  \\
         2458643.96862207  &   -26.28  &  20.7  \\
         2458643.97616583  &   -62.94  &  20.5  \\
         2458643.98371549  &  -118.68  &  20.4  \\
         2458643.99126515  &  -140.81  &  20.9  \\
         2458643.99880301  &  -172.12  &  20.8  \\
         2458644.00634677  &  -211.39  &  23.9  \\
         2458644.01389052  &  -248.11  &  22.6  \\
         2458644.02144019  &  -240.71  &  20.5  \\
         2458644.02898395  &  -247.63  &  22.7  \\
         2458644.03652771  &  -275.91  &  24.0  \\
         2458644.04407147  &  -271.88  &  22.8  \\
         2458644.05161522  &  -243.25  &  23.6  \\
         2458644.05963711  &  -220.53  &  21.5  \\
         2458644.06717497  &  -197.91  &  22.0  \\
         2458644.07471873  &  -170.75  &  21.6  \\
         2458644.08226839  &   -89.29  &  20.0  \\
         2458644.08981215  &   -56.24  &  19.4  \\
         2458644.09736181  &   -15.08  &  17.0  \\
         2458644.10489967  &    -1.27  &  19.4  \\
         2458644.11244343  &    41.11  &  20.4  \\
         2458644.11999309  &   136.84  &  21.5  \\
         2458644.12753094  &   192.76  &  19.5  \\
         \hline

    \end{tabular}
    \caption{Keck/HIRES radial velocities of G 227-22}
    \label{tab:g227-22_hiresrv}
\end{table}

\begin{table}[h!]
    \centering
    \begin{tabular}{| c c c |}
         \hline
         BJD & RV & $\sigma_{\textrm{RM}}$  \\
          & ($\unit{m} \unit{s}^{-1}$) & ($\unit{m} \unit{s}^{-1}$) \\
          \hline
         2458234.9262919190  &   100.71  &  12.0  \\
         2458237.9207755327  &    59.37  &  10.1  \\
         2458238.9732010695  &  -110.79  &  13.1  \\
         2458263.8229116746  &   -43.32  &  13.2  \\
         2458264.8128271060  &    45.73  &   9.8  \\
         2458265.8844489027  &   -67.84  &  10.5  \\
         2458266.8188911910  &   124.41  &   8.7  \\
         2458267.8106331234  &  -157.50  &   8.7  \\
         2458270.8266936536  &   -40.18  &   9.8  \\
         2458288.8428498090  &  -111.37  &   9.7  \\
         2458289.8419703620  &   101.84  &  12.9  \\
         2458291.7777495836  &   131.88  &  10.8  \\
         2458292.8361749600  &    64.94  &  11.1  \\
         2458295.8200156377  &  -157.44  &  10.7  \\
         2458297.8230540957  &   -69.10  &  12.9  \\
         2458299.8125872920  &    -8.08  &  10.8  \\
         2458301.8135485550  &    97.53  &   9.2  \\
         2458319.6796065294  &   -91.42  &   8.7  \\
         2458322.6584219850  &   -17.01  &   9.1  \\
         2458380.5818420276  &   126.82  &   8.0  \\
         2458384.6017408385  &  -138.24  &  11.4  \\
         2458672.8121379544  &    67.75  &  13.4  \\
         \hline

    \end{tabular}
    \caption{HET/HPF radial velocities of G 227-22}
    \label{tab:g227-22_hpfrv}
\end{table}

\begin{table}[h!]
    \centering
    \begin{tabular}{| c c c |}
         \hline
         BJD & RV & $\sigma_{\textrm{RM}}$  \\
          & ($\unit{m} \unit{s}^{-1}$) & ($\unit{m} \unit{s}^{-1}$) \\
          \hline
         2458621.80730122  &   124.30  &  19.4  \\
         2458621.82597166  &   107.93  &  24.3  \\
         2458621.84392196  &    83.32  &  19.5  \\
         2458621.86188407  &    31.49  &  23.7  \\
         2458621.87985209  &   -19.28  &  22.9  \\
         2458621.89779058  &   -74.02  &  20.8  \\
         2458621.91573499  &   -80.56  &  19.4  \\
         2458621.98745946  &  -151.92  &  20.8  \\
         2458622.00542157  &  -154.53  &  21.3  \\
         2458622.02337187  &  -167.00  &  19.7  \\
         2458622.08690632  &  -134.83  &  31.3  \\
         2458643.76038372  &    99.83  &  22.6  \\
         2458643.77832782  &    87.79  &  21.4  \\
         2458643.79627192  &   119.62  &  22.9  \\
         2458643.81438720  &   127.93  &  22.2  \\
         \hline

    \end{tabular}
    \caption{Keck/HIRES radial velocities of GJ 3959}
    \label{tab:gj3959_hiresrv}
\end{table}

\begin{table}[h!]
    \centering
    \begin{tabular}{| c c c |}
         \hline
         BJD & RV & $\sigma_{\textrm{RM}}$  \\
          & ($\unit{m} \unit{s}^{-1}$) & ($\unit{m} \unit{s}^{-1}$) \\
          \hline
         2458502.0311931707  &    -6.24 &  23.5 \\
         2458504.0282843537  &    58.39 &  12.5 \\
         2458507.0036187563  &    45.01 &  17.3 \\
         2458509.0223591705  &   -12.11 &  12.1 \\
         2458519.9991261870  &  -174.57 &  41.4 \\
         2458521.9781113267  &    66.90 &  24.9 \\
         2458523.9701593937  &    74.37 &  15.1 \\
         2458529.9617544464  &   -28.89 &  32.5 \\
         2458531.9664613430  &  -172.92 &  29.2 \\
         2458543.9288727940  &    68.93 &  12.8 \\
         2458621.9416779354  &   -82.86 &  16.4 \\
         2458727.6370346285  &   -28.32 &   6.2 \\

         \hline

    \end{tabular}
    \caption{HET/HPF radial velocities of GJ 3959}
    \label{tab:gj3959_hpfrv}
\end{table}

\end{document}